\documentclass{aa} 

\bibpunct{(}{)}{;}{a}{}{,} 

\usepackage{graphicx}
\usepackage{units}
\usepackage[varg]{txfonts}
\begin{document}

\title{Comparison of the power-2 limb-darkening law from the 
\textsc{Stagger}-grid to Kepler light curves of transiting exoplanets.
\thanks{Tables 1 and 2  only available at the CDS
via anonymous ftp to cdsarc.u-strasbg.fr (130.79.128.5) or via
http://cdsarc.u-strasbg.fr/viz-bin/qcat?J/A+A/000/A00} }

\titlerunning{Power-2 limb-darkening from the \textsc{Stagger}-grid}
\author{P.~F.~L.~Maxted \inst{1} }           
          
\institute{Astrophysics Group,  Keele University, Keele, Staffordshire,
ST5~5BG, UK\\
\email{p.maxted@keele.ac.uk}
}

\date{Dates to be inserted}

 
  \abstract
{Inaccurate limb-darkening models can be a significant source of 
error in the analysis of the light curves for transiting exoplanet and
eclipsing binary star systems, particularly for high-precision light curves at
optical wavelengths. The power-2 limb-darkening law, $I_{\lambda}(\mu) = 1 -
c\left(1-\mu^{\alpha}\right)$, has recently been proposed as a good compromise
between complexity and precision in the treatment of limb-darkening.}
{My aim is to develop a practical implementation of the power-2
limb-darkening law and to quantify the accuracy of this implementation.}
{I have used synthetic spectra based on the 3D stellar atmosphere models from
the \textsc{Stagger}-grid to compute the limb-darkening for
several passbands (UBVRI, CHEOPS, TESS, Kepler, etc.). The parameters of the
power-2 limb-darkening laws are optimized using a least-squares fit to a
simulated light curve computed directly from the tabulated $I_{\lambda}(\mu)$
values. I use the transformed parameters $h_1 =
1-c\left(1-2^{-\alpha}\right)$ and $h_2 = c2^{-\alpha}$ to directly compare
these optimized limb-darkening parameters to the limb darkening measured
from Kepler light curves of 16 transiting exoplanet systems.}
{The posterior probability distributions (PPDs) of the transformed parameters
$h_1$ and  $h_2$ resulting from the light curve analysis are found to be much
less strongly correlated than the PPDs for $c$ and $\alpha$. The agreement
between the computed and observed values of ($h_1$, $h_2$) is generally very
good but there are significant differences between the observed and computed
values for Kepler-17, the only star in the sample that shows significant
variability between the eclipses due to magnetic activity (star spots).  }
{The tabulation of $h_1$ and $h_2$ provided here can be used to accurately
model the light curves of  transiting exoplanets. I also provide estimates of
the  priors that should be applied to transformed  parameters $h_1$ and
$h_2$ based on my analysis of the Kepler light curves of 16 stars with
transiting exoplanets.}

\maketitle
%

\section{Introduction}

 The  specific intensity emitted from a stellar photosphere depends on both
wavelength and viewing angle. For a source function that varies linearly with
optical depth the specific intensity can be described by a linear limb
darkening law, $I_{\lambda}(\mu) = I_\lambda(1)\left[1 - u(1-\mu)\right]$, where
$\mu = \cos(\theta)$ is the cosine of the angle between the line of sight and
the surface normal vector, and the wavelength dependence is implicit in the
linear limb-darkening coefficient, $u$ \citep{Schwarzschild1906}. In general,
limb-darkening is more pronounced at shorter wavelengths, becoming almost
negligible at infrared wavelengths. At optical wavelengths the best-fit
linear limb-darkening coefficient computed from stellar atmosphere models are
$u\approx 0.3$ for hot stars (effective temperature T$_{\rm eff} \ga 15,000$\,K)
and then increases with decreasing T$_{\rm eff}$ from this value to $u\approx
0.9$ for stars with  T$_{\rm eff} \approx 3500$\,K
\citep[e.g.,][]{2004A&A...428.1001C}. Realistic stellar model atmospheres show
that the actual limb darkening of solar-type stars varies by 10\% or more
around the best-fit linear limb-darkening law
\citep[e.g.,][]{2011MNRAS.418.1165H}. 

\begin{figure}
  \includegraphics[width=0.49\textwidth]{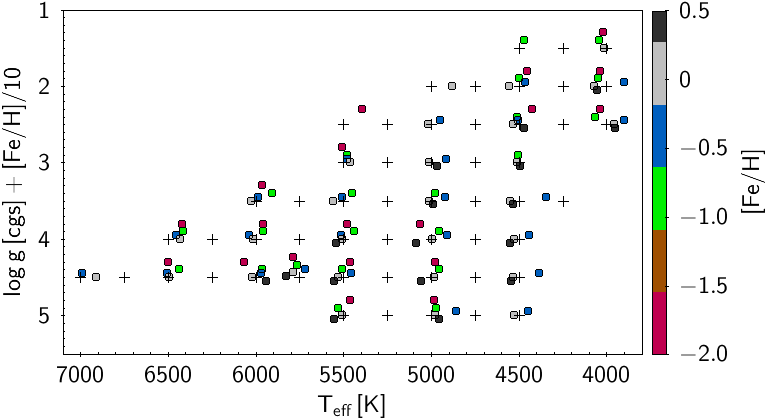}
  \caption{The grid of  \textsc{Stagger}-grid atmosphere models used to
  calculate the limb-darkening profiles of cool stars in this study. Note that
  the points are offset vertically according to [Fe/H] at each value of $\log
  g$ so that they can be distinguished, but no horizontal offset has been
  applied, i.e., the plotted positions give the actual value of T$_{\rm eff}$
  for each model. Small crosses show the regular grid of T$_{\rm eff}$ values
  that have been used for the tabulations provided here. }
  \label{Tlogg}
\end{figure}

 The advent of very high precision photometry for transiting exoplanet systems
has led to extensive discussion in the literature of the systematic errors in
the parameters for these exoplanet systems that result from inaccuracies and
uncertainties in the treatment of limb darkening, e.g.,
\citet{2016MNRAS.457.3573E}, \citet{2013A&A...560A.112M},
\citet{2011MNRAS.418.1165H}, \citet{2008ApJ...686..658S},
\citet{2017AJ....154..111M}, \citet{2017ApJ...845...65N},
\citet{2013MNRAS.435.2152K}, etc.
One well-established result from such studies is that using a linear limb
darkening law can lead to significant bias in the parameters derived from the
analysis of high quality photometry. For example, \citet{2016MNRAS.457.3573E}
found systematic errors in the radius estimates for small planets as large as
3\% as a result of using linear limb-darkening coefficients. There are several
alternative ways to parametrize limb-darkening. Among the alternative
two-parameter laws, the most commonly used in exoplanet studies is the
quadratic limb-darkening law \citep{1950HarCi.454....1K} --
\[ 
I_X(\mu) = 1- c_1(1-\mu) - c_2(1-\mu)^2.
\]
This limb-darkening law has the advantage of being relatively simple and
well-understood in terms of the correlations between the coefficients
\citep{2008MNRAS.390..281P, 2011ApJ...730...50K, 2011MNRAS.418.1165H}	 and how
to sample the parameter space to achieve a non-informative prior
\citep{2013MNRAS.435.2152K} but it fails to match optical high-precision light
curves of transiting exoplanet systems \citep{2007ApJ...655..564K}.
 
  The Claret 4-parameter limb-darkening law \citep{2000A&A...363.1081C} is
often used in exoplanet studies. As the name suggests, this limb-darkening law
uses four coefficients to capture the detailed shape of the limb-darkening
profile $I_X(\mu)$ for a given bandpass $X$ using the following equation -- 
\[ I_X(\mu) = 1 - \sum_{j=1}^{4}a_j(1-\mu^{j/2}).\] 
Including the coefficients of this limb-darkening law in a least-squares fit
to observed transit light curves is impractical because they are found to be
strongly correlated with each other and degenerate with other parameters of
the fit. Instead, it is common practice to use interpolation within a grid
of tabulated coefficients \citep[e.g.,][]{2013A&A...552A..16C} to select the
values of $a_1$,\,\dots,\,$a_4$. These coefficients may be fixed or the
effective temperature used for the interpolation, $T_{\rm eff,ld}$, can be
included as a free parameter in the fit \citep[e.g.,][]{2016A&A...591A..55M}.

Among the limb-darkening laws with 2 coefficients, the power-2 limb-darkening
law  \citep{1997A&A...327..199H} has been recommended by
\citet{2017AJ....154..111M} as they find that it outperforms other
two-coefficient laws adopted in the exoplanet literature in most cases, 
particularly for cool stars. The form of this limb-darkening law is
\[I_X(\mu) = 1-c\left(1-\mu^{\alpha}\right).\]
Using  an exponent of $\mu$ rather than a coefficient of some function of
$\mu$ enables this two-parameter law to match accurately the shape of the limb
darkening profile towards the limb of the star using only one extra parameter
cf. a linear limb-darkening law. There has been very little discussion in the
literature of the power-2 limb-darkening law applied to exoplanet transit
studies and so its properties and the practicalities of using this law to
determine exoplanet properties are not yet well understood. Here I describe a
practical implementation of the power-2 limb-darkening law, including a
tabulation of the parameters for various passbands and instruments, and
present  an analysis of the Kepler light curves for 16 transiting exoplanet
systems that has been used quantify the accuracy of these parameters.

\section{Analysis}

 The following section describes the calculation of the limb-darkening
profiles, $I_X(\mu)$, for various passbands and how the parameters of a
power-2 limb-darkening law based on these profiles can be optimized for the
analysis of the light curve for a given exoplanet system or binary star. I
then compare these optimized power-2 limb-darkening law parameters to
observed values for transiting planet host stars derived from the analysis of
their Kepler light curves. This comparison shows a very good level of
agreement between theory and observations so I then proceed to provide a
tabulation of power-2 limb-darkening law parameters that can be used to
model the light curves of exoplanet systems and binary stars together with
some recommendations for their use.

\subsection{Calculation of the limb-darkening profiles \label{ldsect}}
 I have calculated the variation of specific intensity with viewing angle
integrated over various passbands, i.e., the limb-darkening profile,
$I_X(\mu)$, where the subscript $X$ denotes the passband. For the specific
intensity as a function of wavelength and viewing angle, $I_{\lambda}(\mu)$, I
have used the synthetic 3D LTE spectra from the \textsc{Stagger}-grid
calculated by \citet{2015A&A...573A..90M}. This grid of models spans a
range of effective temperature, surface gravity and metallicity that covers
the majority of known exoplanet host stars. There are small but, for
high-precision work, significant differences between the limb
darkening predictions from 3D model atmospheres compared to 1D model atmospheres
\citep{2015A&A...573A..90M}. 3D model atmospheres provide a more realistic
description of the convective motions at different heights in cool star
atmospheres than the simplified mixing-length approach adopted in 1D stellar
model atmospheres. This is essential for accurate prediction of the limb
darkening, which is determined mainly by the temperature gradient. The limb
darkening predictions from 3D model atmospheres are a good match  to the
observed limb darkening profiles of the Sun \citep{2013A&A...554A.118P},
$\alpha$~Cen A and B \citep{2006A&A...446..635B} and HD~209458
\citep{2012A&A...539A.102H}. In all three cases, the 3D model atmospheres
provide a better match to the observations than 1D model atmospheres. The 3D
radiative hydrodynamic atmosphere models from which  the spectra used
here are calculated are described in more detail by
\citet{2018arXiv180101895C}. The grid of stellar atmosphere models covers the
effective temperature  range T$_{\rm eff} \approx 4000$\,K to 7000\,K in steps
of approximately 500\,K, the surface gravity range $\log g = 1.5$ to 5.0 in
steps of 0.5 dex, and metallicity values of [Fe/H]$ = -2.0$, $-1.0$, $-0.5$,
0.0 and $+0.5$. The coverage within these ranges is not uniform -- the
complete set of available models is shown in Fig.~\ref{Tlogg}. The effective
temperature assigned to each model using the Stefan-Boltzmann law applied to
the computed emergent spectrum is taken from Table~B0 of
\citet{2018A&A...611A..11C}. The spectra computed from the
\textsc{Stagger}-grid model atmospheres are provided at 11 values of $\mu$
from 0 to 1 for [Fe/H]$ = -2, -1$ and 0, and at 10 values of $\mu$ from 0.01
to 1 for the models at [Fe/H]$ = \pm 0.5$. The computed spectra for $\mu=0$
have very low flux levels compared to the other spectra so for [Fe/H]$ = \pm
0.5$ I assume $I_X(0) = 0$. The spectra are sampled with variable wavelength
step $\Delta\lambda$ such that $\lambda/\Delta\lambda = 20\,000$.

 The limb-darkening profile, $I_X(\mu)$, has been calculated for the following
instruments and  passbands:
CHEOPS \citep{2017SPIE10563E..1LC};
Kepler  \citep{2010Sci...327..977B}; 
TESS \citep{2015JATIS...1a4003R};
Gaia \citep{2016A&A...595A...1G};
Johnson/Bessell UBVRI \citep{1990PASP..102.1181B};
Sloan SDSS $ugriz$ \citep{2010AJ....139.1628D};
MOST \citep{2003PASP..115.1023W};
CoRoT \citep{2009A&A...506..411A}.
 
 The spectra at [Fe/H]$ = \pm 0.5$ only cover the wavelength range up to
1\,$\mu$m. The bandpasses for the CHEOPS and TESS instruments extend a small
way beyond this limit. This was handled by calculating a linear fit to the log
of the model flux, $\log f_{\lambda}$ over the region 0.8\,$\mu$m to
1\,$\mu$m, and then extrapolating this linear fit up to 1.15\,$\mu$m.
Comparing this linear extrapolation to the computed flux at other values of
[Fe/H] shows that it provides a very good match to the actual flux
distribution in this region, on average. In addition, the fraction of the
stellar flux emitted in this extrapolated wavelength region for the CHEOPS and
TESS bandpasses is $\la1\%$ so this extrapolation will introduce negligible
systematic error in the computed values of $I_X(\mu)$. 

	The nature of the detector used in the instrument must be accounted for in
the calculation of the limb-darkening profile \citep{1998A+A...333..231B}.
Charge-coupled devices (CCDs) count photons, they do not measure energy, so for
the accurate computation of $I_X(\mu)$ over a passband with response function
$R_X(\lambda)$ for an instrument with a CCD detector,  the specfic intensity
must be converted from energy flux to photon number flux, i.e.,
\[ I_X(\mu) = \int_0^{\infty} R_X(\lambda) I_{\lambda}(\mu)/ 
(hc/\lambda)\,d\lambda.\] 
This has little effect for narrow passbands but can be a noticable effect for
the broad-band photometers often used for planetary transit surveys such as
Kepler and CHEOPS.

 The irregular spacing of T$_{\rm eff}$ values within the model grid and their
relatively large spacing of about 500\,K can be difficult to deal with when
interpolating within the grid of $I_X(\mu)$ values. To remedy this problem I
have used interpolation of $I_X(\mu)$ as a function T$_{\rm eff}$ 
at each value of $\log g$, [Fe/H] and $\mu$ to create a tabulation of
$I_X(\mu)$ on a regular grid of  T$_{\rm eff}$ with a grid spacing of 250\,K.
This requires a small degree of extrapolation at the edge of the model grid so
I used quadratic spline interpolation to ensure that this extrapolation is
stable. In a few cases there are only two values of T$_{\rm eff}$ for a given
pair of $\log g$ and [Fe/H] values, in which case I use a linear fit to
$I_X(\mu)$ points for the interpolation and extrapolation. The model spectra
for $\log g =4.44$ and $(\log g, [{\rm Fe/H}]) = (3.0, -2)$ are only available
for one value of  T$_{\rm eff}$ so there are no corresponding entries for
these $\log g$ and [Fe/H] values in the tabulations provided here. A complete
tabulation of $I_X(\mu)$ for each passband is provided in the electronic
version of this journal, an excerpt from this table is shown in
Table~\ref{ImuTable}.

\begin{figure}
  \includegraphics[width=0.45\textwidth]{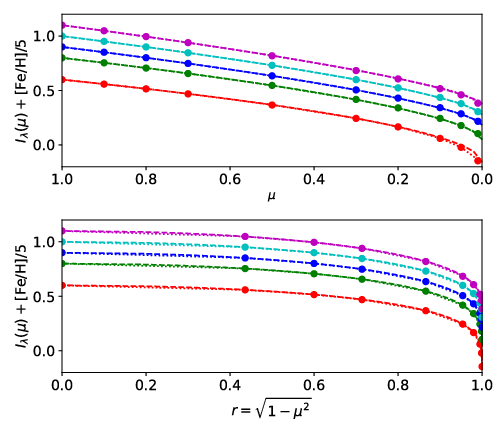}
  \caption{Comparison of directly calculated (points) and interpolated (dotted
  lines) limb-darkening profiles for the CHEOPS bandpass at T$_{\rm eff}
  \approx 5777$\,K, $\log g = 4.44$ and for [Fe/H]$ = -2, -1, -0.5,  0, +0.5$.
  The corresponding power-2 limb-darkening laws optimized to fit the light
  curve for a typical Jupiter-like transiting exoplanet are also shown (dotted
  lines).  Note that the profiles are offset vertically according to [Fe/H].
  The lower panel is plotted as a function of the distance from the centre of
  the stellar disc in units of the stellar radius.}
  \label{TestMuGridInterpolator}
\end{figure}

The limb-darkening profiles for T$_{\rm eff}\approx 5777$\,K are not included
in tabulation of $I_X(\mu)$ so interpolating in this table to match these
profiles is a good test of the tabulated values and the interpolation scheme
used.   In Fig.~\ref{TestMuGridInterpolator} we compare the directly computed
values of $I_X(\mu)$ for T$_{\rm eff}\approx 5777$\,K and $\log g =4.44$ to
the interpolated values using linear barycentric
interpolation.\footnote{implemented in the {\sc python} module
\texttt{scipy.interpolate} as the class \texttt{LinearNDInterpolator}} The rms
residual between the computed and interpolated values is 0.5\% or less over
the $\mu$ range 0.01 to 1.0 for these data. Also shown in
Fig.~\ref{TestMuGridInterpolator} are  the corresponding power-2
limb-darkening laws where the parameters have been optimized as described
Section~\ref{p2sect}.

\begin{table*}
\caption{Specific intensity as function of the cosine of the viewing angle
  integrated over various bandpbasses, $I_X(\mu)$. The bandpass, $X$, is noted
  in the first column as either the filter name for the Johnson photometric
  system, `u\_', `g\_', etc. for the SDSS photometric system, or the two
  initial letters for Kepler, CHEOPS, etc. This is an excerpt from the
  complete table that is available at the CDS, provided here as a guide to its
  format and content.}
\label{ImuTable}
\begin{center}
\begin{tabular}{@{}lrrrrrrrrrrrrrr}
\hline
$X$&\multicolumn{1}{l}{T$_{\rm eff}$} &
\multicolumn{1}{l}{$\log g$} &
\multicolumn{1}{l}{[Fe/H]} &
\multicolumn{9}{c}{$\mu$} \\
&\multicolumn{1}{l}{[K]} &
\multicolumn{1}{l}{[cgs]} &
 & 0 & 0.01 & 0.05 & 0.10 & 0.20 & 0.30 & 0.50 & 0.70 & 0.80 & 0.90 &1\\
\hline
\noalign{\smallskip}
Ke&4000&1.5&-1.0& 0& 0.2088& 0.2933& 0.3476& 0.4312& 0.5067& 0.6519& 0.7937& 0.8633& 0.9321& 1\\
Ke&4250&1.5&-1.0& 0& 0.2397& 0.3192& 0.3738& 0.4585& 0.5346& 0.6767& 0.8106& 0.8750& 0.9381& 1\\
Ke&4500&1.5&-1.0& 0& 0.2705& 0.3450& 0.4000& 0.4858& 0.5625& 0.7015& 0.8274& 0.8867& 0.9442& 1\\
Ke&4000&2.0&-2.0& 0& 0.1926& 0.2763& 0.3385& 0.4318& 0.5125& 0.6613& 0.8017& 0.8693& 0.9354& 1\\
Ke&4250&2.0&-2.0& 0& 0.2167& 0.3029& 0.3665& 0.4629& 0.5450& 0.6901& 0.8211& 0.8828& 0.9424& 1\\

\noalign{\smallskip}
\hline
\end{tabular}
\end{center}
\end{table*}

\subsection{Kepler light curves}
 To test the accuracy of the power-2 limb-darkening profiles calculated from
the {\sc Stagger}-grid I have compared model light curves generated from these
data to high-precision light curves for transiting exoplanet systems observed
with Kepler. Transiting exoplanet systems with deep eclipses observed at high
signal-to-noise with Kepler were selected from the list of 38 systems studied
by \citet{2013A&A...560A.112M}. I excluded stars for which there is no
published estimate of T$_{\rm eff}$, $\log g$ and [Fe/H] based on
high-resolution spectroscopy and also systems with stellar companions or
eccentric orbits. There are also a few stars that had to be excluded from
this study because they are too hot or too cool, i.e., the observed values of
T$_{\rm eff}$, $\log g$  and [Fe/H] for these stars lie outside the grid of
limb-darkening profiles from the {\sc Stagger}-grid.
 
 Short cadence (SC) light curves from Data Release 25
\citep[DR25,][]{2016ksci.rept....9T} for the remaining 16 systems were
downloaded from the the Mikulski Archive for Space Telescopes\footnote{\url
{https://archive.stsci.edu/kepler/}} (MAST). DR25 is the final data release
from the Kepler mission and contains all the observation obtain during the
original Kepler mission divided in 18 ``quarters'' (Q0 to Q17) with a typical
duration of about 90 days  each. I used the observations provided in the
column PDCSAP\_FLUX and their associated errors for this analysis.
Observations with a non-zero value of the SAP\_QUALITY quality flag were
excluded from further processing. I also excluded outliers that deviated by
more than 4 standard deviations from a running median filter. Trends in the
data due to instrumental and stellar noise sources were modelled using a
Gaussian process (GP) fitted to the data between the transits. I used the {\sc
celerite} package \citep{2017AJ....154..220F} to model these data using the
following kernel with the default value of $\epsilon=0.01$ to  approximate the
Mat\'{e}rn-3/2 covariance function: 
\[ k(\tau) = \sigma^2\,\left[
  \left(1+1/\epsilon\right)\,e^{-(1-\epsilon)\sqrt{3}\,\tau/\rho}
  \left(1-1/\epsilon\right)\,e^{-(1+\epsilon)\sqrt{3}\,\tau/\rho} \right]. \]
 Here, $\tau$ is the time difference between two observations and $\rho$
is a parameter that controls the time scale over which observational errors
are correlated. Evaluating the GP at all data points was slow so instead it
was evaluated at 1000 points evenly distributed in time across each quarter of
the Kepler data and then spline interpolation was used to evaulate the trend
for all the observations in each quarter, including those obtained during a
transit. The data were divided by the trend and saved with their time of
observation and errors for further analysis.

\subsection{Light curve analysis}

 I used version 1.8.0 of the \texttt{ellc} light curve model
\citep{2016A&A...591A.111M} to determine the geometry and limb-darkening
parameters for each of these 16 transiting exoplanet systems. The free
parameters in the model for each
system were: the radius of the host star in units of the semi-major axis, 
$R_{\star}/a$, the ratio of the radii,  $k = R_{pl}/R_{\star}$, where $R_{\rm
pl}$ is the radius of the companion; the impact parameter,  $b = a
\cos(i)/R_{\star}$, where $i$ is the  orbital inclination; the time of primary
eclipse,  $T_0$; the orbital period,  $P$; and the power-2 limb-darkening
parameters,  $c$ and $\alpha$. For systems where contamination of the
photometric aperture is noted in the MAST archive data I also include a
``third-light'' parameter, $\ell_3$, as a free parameter in the fit with a prior
set by the mean and standard deviation of the contamination values given for
each Kepler quarter. These contamination values are all very small ($\la 1$\%)
so they have little influence on the analysis.
 
 I assume that the star in these exoplanet systems is spherical and  use a
polytrope with index $n=1.5$ to calculate the dimensions of the ellipsoid used
to approximate the shape of the planet. The flux contribution in the Kepler
bandpass from the planet is assumed to be negligible for these systems. There
is little or no information about the geometry of these systems in the
observations between the eclipses so this analysis  uses only observations
obtained during transit plus an additional 25\% of  the transit width before
and after start and end of the transit. Circular orbits have been assumed for
all systems.

 Some of the targets studied here show quasi-periodic variability due to
magnetic activity (star spots). It is notoriously difficult to include star
spots in the model for a transiting planet system because the number of free
parameters required is large and the constraints on these parameters from the
light curve are generally weak and highly degenerate. I did not attempt to
model star spots for any of the systems here. Instead, I  simply divide-out
the trend  established from the Gaussian process fit to the out-of-eclipse
data. The effect of this approach on the results will be discussed below.

I used {\sc emcee} \citep{2013PASP..125..306F}, a {\sc python} implementation
of an affine invariant Markov chain Monte Carlo (MCMC) ensemble sampler, to
calculate the posterior probability distribution (PPD) of the model
parameters. An ensemble with 72 or (for cases where $\ell_3$ was a free
parameter) 80 samples per chain step (``walkers'') was initialized using 512
``burn-in'' chain steps. The PPD was then calculated using 256 chain steps
starting with the last ensemble from the burn-in phase. The convergence of the
chain was judged by visual inspection of the parameters and the likelihood as
a function of step number. In cases where it was suspected that the chain had
not sampled the posterior probability distribution accurately, a new Markov
Chain was calculated starting from the best-fit parameters in the previous
chain. This initial optimization of the parameters was done using the
``grid\_1 = very\_sparse'' option in \texttt{ellc} to specify the minimum
number of grid points for the numerical integration of the flux from the host
star. This has the advantage of reducing the computation time but introduces a
bias of approximately 50\,ppm in the simulated eclipse depth. This bias is
significant for the analysis of the very high precision light curves being
used in this study so the results below  were calculated using a second Markov
chain with 256 chain steps following a burn-in phase of 128 steps using the
option ``grid\_1 = default''. The second chain was initialised using the best
fit parameters found from the first Markov chain. In practice, this made a
neglible difference to the result presented below for $h_1$ and $h_2$ but
there are very small but significant differences to the parameters
$R_{\star}/a$, $k$ and $b$ for some stars.

The standard error for observation $i$ was assumed to be $\sigma_{{\rm
tot},i}^2 = \sigma_{\rm ext}^2 + \sigma_{i}^2$ where $\sigma_{i}$ is the
standard error derived from the value PDCSAP\_FLUX\_ERR in the MAST archive
file and $\sigma_{\rm ext}$ is a constant that is optimized by including the
relevant term in the calculation of the likelihood for each trial set of
parameters. For most of the light curves studied here we find $\sigma_{\rm
ext}\approx 20$\,ppm.  If the out-of-eclipse level is included as a free
parameter it is found to be always very close to the value 1 with a very small
error and is not correlated with the other parameters so it was fixed this
value for the analysis presented here. 

A summary of the results from this analysis are given in
Table~\ref{ResultsTable}. The best fit found for each light curve is shown in
Fig.~\ref{lcfitFig}. The values of $T_0$ and $P$ are not quoted here because
the long cadence (LC) data from Kepler and other times of mid-eclipse reported
in the literature are not included in the analysis and so these estimates are
not optimal. We also do not quote directly the values of $c$ and $\alpha$
determined from these fits for the reasons described in the following
section.

\subsection{Comparison of observations and theory}
The joint PPDs for $c$ and $\alpha$ calculated using {\sc emcee} from the
Kepler light curves are shown in Fig.~\ref{h1h2Fig}. It can be seen that these
parameters are strongly correlated, i.e., neither parameter is determined very
accurately,  but the shape of the transit light curve does impose a strong
constraint on the relationship between these two parameters. This makes it
awkward to compare the computed values of these parameters directly to the
observed values because the two observed values may agree within their error
bars with the computed values while being inconsistent with the constraints on
the relationship between them imposed by the light curve.

Instead of $c$ and $\alpha$, I use the parameters 
\begin{equation}
  h_1 = I_X(\nicefrac{1}{2}) = 1-c\left(1-2^{-\alpha}\right)
\end{equation}
and
\begin{equation}
  h_2 = I_X(\nicefrac{1}{2}) - I_X(0) = c2^{-\alpha}
\end{equation}
to compare the computed and observed limb-darkening profiles.  The value of
$h_1$ measures the bandpass-integrated specific intensity relative to the
centre of the disc, $I_X$, in the region on the stellar disc at a distance
$r=\sqrt{1-\left(\nicefrac{1}{2}\right)^2}\approx 86.6\%$ towards the limb. Similarly, $h_2$
measures the drop in $I_X$ between the same radius and the limb. These
definitions impose the constraints $h_1 < 1$ and $h_1+h_2 \le 1$ that are
required so that the flux is positive at all points on the stellar disc. The
inverse relations, provided here for convenience, are
\begin{equation}
 c = 1 - h_1 + h_2 
\end{equation}
and
\begin{equation}
 \alpha = \log_2 \left([1 - h_1 + h_2]/h_2\right) = \log_2(c/h_2).
\end{equation}

\begin{figure}
\includegraphics[width=0.49\textwidth]{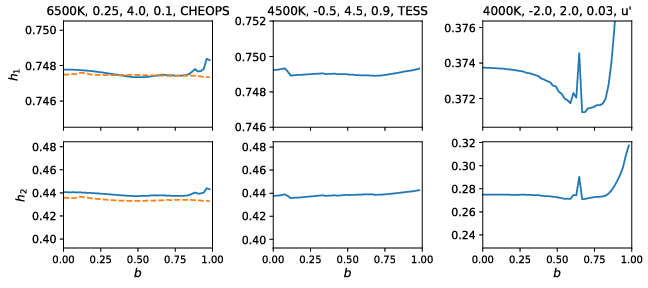}
\caption{Optimized and transformed power-2 limb-darkening law parameters,
$h_1$ and $h_2$ as a function of impact parameter, $b$. The values of T$_{\rm
eff}$, [Fe/H], $\log g$ and radius ratio, $k$, and the bandpass used for the
simulation are noted in the title to each pair of panels. Note that the
vertical scale on each axis is set by the median standard errors on $h_{1,{\rm
obs}}$ and $h_{2,{\rm obs}}$ from Table~\protect\ref{ResultsTable} ($\pm 0.003$
and $\pm 0.046$, respectively). Also shown for the CHEOPS passband are the
results for $k=0.9$ (dashed line). The small spikes seen on the curves for the
$u^{\prime}$ passband are the result of numerical noise in the light curves at
the few ppm level. }
\label{PlotPower2} 
\end{figure}

The values of $h_1$ and $h_2$ derived from the analysis of the Kepler light
curves are given in Table~\ref{ResultsTable}. Also given in this table are the
optimized values of $h_1$ and $h_2$ computed from the \textsc{Stagger}-grid,
as described below.

\begin{figure}
  \includegraphics[width=0.49\textwidth]{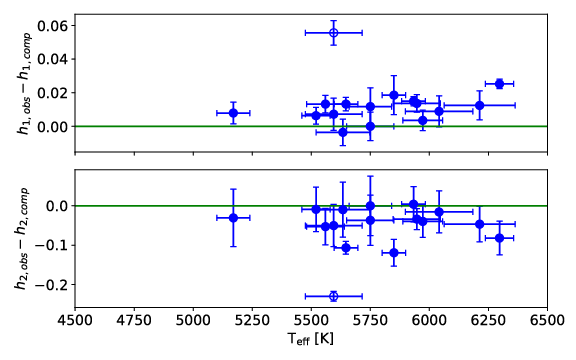}
  \caption{Difference between the observed and computed values of $h_1$ and 
  $h_2$. Data for Kepler-17 are plotted using an open symbol. The vertical
  error bars are calculated from the square root of the sum of the variances
  of the observed and calculated values. }
  \label{h1h2Teff}
\end{figure}
\subsection{Optimized power-2 limb-darkening law parameters \label{p2sect}}

\citet{2011MNRAS.418.1165H} has outlined a simple and elegant method to
achieve a ``like-for-like'' comparison between observed and theoretical
limb-darkening profiles for transiting exoplanet systems. The essence of the
method is to simulate light curves using the limb-darkening profile and then
to find the parameters of the limb-darkening law that provide the
best least-squares fit to this simulated light curve. These optimized
limb-darkening parameters can then be compared directly to the values obtained
by fitting the light curves of transiting exoplanet systems.

I used version 1.8.0 of the \texttt{ellc} light curve model to simulate light
curves of transiting exoplanet systems using the limb-darkening profile data
for the Kepler bandpass described in Section~\ref{ldsect}. After some
experimentation I found that a simulated light curve with 32 points uniformly
distributed from mid-transit to the end of the transit is sufficient to
calculate $h_1$ and $h_2$ to a precision much better than the observed
accuracy of these parameters. Similarly, the grid size option ``very\_sparse''
is sufficient to precisely calculate the $h_1$ and $h_2$ for the transiting
exoplanet systems studied here, although for the direct comparison with the
values of $h_1$ and $h_2$ from the Kepler light curves in
Table~\ref{ResultsTable} we used the ``default'' grid size option in order to
minimize the numerical noise in these results. The optimization of the
limb-darkening parameters uses spheres to model both the star and the planet
and a tabulation of the limb-darkening profile interpolated using a monotonic
piecewise cubic Hermite interpolating polynomial\footnote{implemented in the
{\sc python} module \texttt{scipy.interpolate} as the class
\texttt{PchipInterpolator}} onto a regular grid of 51 $\mu$ values. The light
curve is independent of the assumed value for $R_{\star}/a =0.1$ with these
assumptions. The values of $b = a\cos{i}/R_{\star}$ and $k=R_{\rm
pl}/R_{\star}$ used for each system were taken from Table~\ref{ResultsTable}.
These calculations were done using the {\sc python} module
\texttt{pycheops}\footnote{\url{https://pypi.python.org/pypi/pycheops/}} that is
currently under development in support of the ESA CHEOPS mission.

The values of  $h_1$ and $h_2$ each define a relation between $c$ and
$\alpha$, as follows --
\begin{equation}
 c = h_2 2^{\alpha} ,
\end{equation}
\begin{equation}
 c = (1-h_1)/(1-2^{-\alpha}).
\end{equation}
These relations are shown in Fig.~\ref{h1h2Fig} for the values of
$h_{1,{\rm comp}}$ and $h_{2,{\rm comp}}$ optimized as described above using
limb-darkening profiles interpolated from Table~\ref{ImuTable}. Unless
otherwise stated, the values of T$_{\rm eff}$, $\log g$ and [Fe/H]  in this
table used to compute $h_{1,{\rm comp}}$ and $h_{2,{\rm comp}}$ were taken
from TEPCat\footnote{\url{http://www.astro.keele.ac.uk/jkt/tepcat}}
\citep{2011MNRAS.417.2166S} using values published up to 2018 Feb 25. The
standard errors on $h_{1,{\rm comp}}$ and $h_{2,{\rm comp}}$ due to the
uncertainties in the observed values of T$_{\rm eff}$, $\log g$ and [Fe/H]
were calculated using a Monte Carlo method, i.e., we generated a small sample
of $h_{1,{\rm comp}}$ and $h_{2,{\rm comp}}$ values using values of T$_{\rm
eff}$, $\log g$ and [Fe/H] randomly selected from Gaussian distributions with
mean and standard deviation set from the observed values with their quoted
standard errors, then took the standard deviations of these samples as the
standard errors for $h_{1,{\rm comp}}$ and $h_{2,{\rm comp}}$. 

\begin{table}
\caption{Optimized power-2 limb-darkening law parameters, $c$ and $\alpha$,
and the corresponding transformed parameters,  $h_1$ and $h_2$. The bandpass,
$X$, is noted in the first column as either the filter name for the Johnson
photometric system, `u\_', `g\_', etc. for the SDSS photometric system, or the
two initial letters for Kepler, CHEOPS, etc. For stars that do not show strong
magnetic activity it can be assumed that $h_1$ is accurate to $\pm \sigma_1
= \pm 0.011$ and $h_2$ is accurate to  $\pm \sigma_2 = \pm0.045$. This is an
excerpt from the complete table that is available at the CDS, provided here
as a guide to its format and content.}
\label{Power2Table}
\begin{center}
\begin{tabular}{@{}lrrrrrrr}
\hline
$X$&\multicolumn{1}{l}{T$_{\rm eff}$ } &
\multicolumn{1}{l}{$\log g$} &
\multicolumn{1}{l}{[Fe/H]} &
  \multicolumn{1}{l}{$c$} &
  \multicolumn{1}{l}{$\alpha$} &
  \multicolumn{1}{l}{$h_1$} &
  \multicolumn{1}{l}{$h_2$} \\
&\multicolumn{1}{l}{[K]} &
\multicolumn{1}{l}{[cgs]} &
& & & & \\
\hline
\noalign{\smallskip}
CH & 4000 &  1.5 & $-1.0$&   0.727 &   0.916 &   0.658 & 0.39\\
CH & 4250 &  1.5 & $-1.0$&   0.722 &   0.841 &   0.681 & 0.40\\
CH & 4500 &  1.5 & $-1.0$&   0.724 &   0.758 &   0.704 & 0.43\\
CH & 4000 &  2.0 & $-2.0$&   0.759 &   0.833 &   0.667 & 0.43\\
CH & 4250 &  2.0 & $-2.0$&   0.764 &   0.739 &   0.694 & 0.46\\
CH & 4500 &  2.0 & $-2.0$&   0.786 &   0.634 &   0.720 & 0.51\\
CH & 4000 &  2.0 & $-1.0$&   0.714 &   0.953 &   0.655 & 0.37\\
CH & 4250 &  2.0 & $-1.0$&   0.717 &   0.880 &   0.673 & 0.39\\
CH & 4500 &  2.0 & $-1.0$&   0.726 &   0.804 &   0.690 & 0.42\\
CH & 4000 &  2.0 & $-0.5$&   0.693 &   0.911 &   0.676 & 0.37\\
  \noalign{\smallskip}
\hline
\end{tabular}
\end{center}
\end{table}

 Fig.~\ref{PlotPower2} shows the how the parameters $h_1$ and $h_2$ vary as a 
function of impact parameter, $b$, for three different test cases. The first
case is similar to the exoplanet systems studied above but assuming that the
observations have been done with CHEOPS rather than Kepler. The second case
simulates an eclipsing binary containing two similar cool dwarf stars observed
with TESS. The third case is a ``worst-case scenario'' of an eclipse of a
metal-poor red giant observed in the Sloan $u^{\prime}$ passband. The vertical
scale in these diagrams is set by the median standard errors on $h_{1,{\rm
obs}}$ and $h_{2,{\rm obs}}$ from Table~\protect\ref{ResultsTable} ($\pm 0.003$
and $\pm 0.046$, respectively). For the first two cases, the variation of
$h_1$ and $h_2$ is always much less than the typical uncertainty on these
values that can be achieved with the best data currently available. The
variations in $h_1$ and $h_2$ are also less than $\pm 0.003$ and $\pm 0.046$
for $b \la 0.9$ in the worst-case scenario. For the first case we also
simulated the case $k=0.9$. In this case, $h_1$ and $h_2$ are also seen
to have negligible dependence  on $k$.

 The optimized parameters of a power-2 limb-darkening law for all passbands
are given in Table~\ref{Power2Table} for all values of T$_{\rm eff}$, $\log
g$ and [Fe/H] in our re-sampled model grid. The optimization has only been
done for the case $b=0$ and $k=0.1$ since, in general, the parameters $h_1$
and $h_2$ show very little dependence on these parameters. The calculation
was done using the grid size option ``default'' in {\sc ellc} to simulate  32
points evenly distributed through one half of a symmetrical transit light
curve, as described above.

\section{Discussion}

 The differences $\Delta h_1 = h_{1,{\rm obs}} - h_{1{\rm comp}}$  and $\Delta
h_2 = h_{2,{\rm obs}} - h_{2{\rm comp}}$ between the observed and computed
values of $h_1$ and $h_2$ are shown as a function of effective temperature,
T$_{\rm eff}$, in Fig.~\ref{h1h2Teff}.  Kepler-17 is a clear outlier in these
plots. This is very likely to be a consequence of the magnetic activity that
is obvious in the light curve of this star. The root mean square of this
variation is about 0.8\% on average in the Kepler SC light curves of
Kepler-17, which is an order of magnitude larger than the next most variable
star and approximately 60 times larger than the median value of this statistic
for the other 15 stars in this sample. For this reason, we ignore Kepler-17 in
the following discussion of the random and systematic errors in $\Delta h_1$
and $\Delta h_2$.

\begin{figure}
  \includegraphics[width=0.49\textwidth]{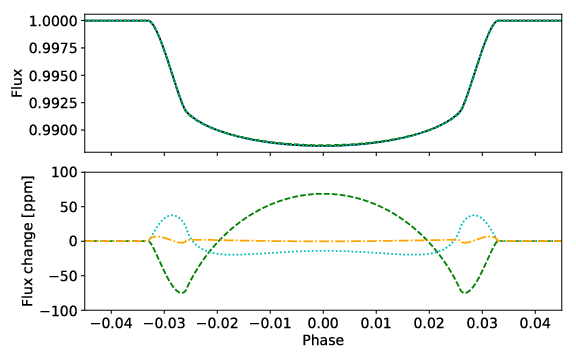}
  \caption{A simulated light curve for the following parameters:
  $R_{\star}$/a=0.2, k=0.1, b=0.4, q=0.0008, $h_1 = 0.75 \pm 0.01$, $h_2 =
  0.45 \pm 0.05$. The upper plot shows the light curve for the nominal values
  of $h_1$ and $h_2$ and (indistinguishable at this scale) the light
  curves for both $h_1$ and $h_2$ perturbed upwards by one standard deviation.
  The lower plot shows the change in flux due to a change by +1 standard
  deviation in $h_1$ (dashed line) and $h_2$ (dotted line). Also shown in the
  lower panel is the effect of using a sphere to model the shape of the
  transiting planet instead of a polytrope (dot-dashed line). Note that in the
  latter case the radius of the sphere has been reduced by 0.5\% cf. the
  volume-average radius of the ellipsoid used to model the planet as a
  polytrope. This correction is required so that the eclipse depth for the
  spherical planet case  matches the eclipse depth for the ellipsoidal planet
  model.}
  \label{h1h2lc}
\end{figure}

The square root of the mean residuals excluding  Kepler-17 are $0.012$ for
$h_1$ and 0.055 for $h_2$. The mean value of the residuals  are $\langle
\Delta h_1 \rangle = 0.010 \pm 0.002$ and $\langle \Delta h_2 \rangle = -0.042
\pm  0.010$, where the uncertainty quoted here is the standard error on the
mean.  If we assume that there is some additional error $\sigma_{1}$ in either
the computed or observed values of  $h_1$, and similarly for $h_2$, then we
find that $\sigma_{1}=0.011$  and $\sigma_{2}=0.045$ are required to achieve
$\chi^2 = N_{\rm df}$ for the null hypothesis $\Delta h_1 = \Delta h_2 = 0$,
where $N_{\rm df} = 15$ is the number of degrees of freedom, which is equal to
the number of observations here since there are no free parameters in this
model.

The origin of this small additional uncertainty in $h_1$, and  $h_2$ will be
discussed below, but it should be stated immediately that these statistics
represent a remarkably good level of agreement between the state-of-the-art in
stellar atmospheric models and the best available observational data for
planet host stars. They also demonstrate that the power-2 limb-darkening law
does indeed represent a very good approximation to both the observed and
theoretical limb-darkening profiles of cool stars. It also has the advantage
of being a simple function of two parameters, and the transformed parameters
$h_1$ and $h_2$ provide a way to represent this limb-darkening law that allows
for direct comparison between theory and observations. These transformed
parameters are have the useful feature that they are not strongly correlated
when used as free parameters for the least-squares fit to the light curves of
transiting exoplanet systems.

 In general, there remains a weak correlation between $h_1$ and $h_2$ in the
PPDs shown in Fig.~\ref{h1h2Fig}. It is possible to reduce or
remove this correlation by adjusting the choice of a reference value of
$\mu_{\rm ref}$ in the definition of the parameters $h_1^{\prime} =
I_X(\mu_{\rm ref}) = 1-c\left(1-\mu_{\rm ref}^{\alpha}\right)$ and
$h_2^{\prime} = I_X(\mu_{\rm ref}) - I_X(0) = c\mu_{\rm ref}^{\alpha}$. We
have not made this adjustment here because the aim is to compare the values of
$h_1$ and $h_2$ for different stars. It is not straightforward to compare the
values of $h_1^{\prime}$ and $h_2^{\prime}$ across a sample of stars because
these values will depend on the different values of $\mu_{\rm ref}$  used for
each transiting exoplanet as well as the properties of the star itself.

Kepler-17 is a clear outlier in terms of its variability between the transits
and also an outlier in Fig.~\ref{h1h2Teff}. A likely reason for this
disagreement can be seen from careful inspection of the residuals from the
light curve fit in Fig.~\ref{lcfitFig}. There is clear excess noise during the
transit that is not seen before or after the transit. This excess noise is due
to the planet transiting magnetically active regions on the star (star spots
and faculae). This strongly suggests that the reason for the offset between
the observed and computed values of $h_1$ and $h_2$ is due to the relatively
high level of magnetic activity in this star. It is also noticable that the
small but significant mean offset between the observed and computed values of
$h_1$ and $h_2$ for the other 15 stars in the sample are both in the same
sense as for Kepler-17, i.e., $\Delta h_1$ positive and  is  $\Delta h_2$ is
negative. This suggests that part of the reason for this offset may be weak
magnetic activity in these solar-type stars that is not  included in the
stellar atmosphere models used here. 

Fig.~\ref{h1h2lc} shows the effect on a typical light curve of perturbing
$h_1$ by $\sigma_1$ and similarly for  $h_2$ and  $\sigma_2$. It can be seen
that $h_1$ influences the overall shape and depth of the transit, whereas the
influence of  $h_2$ is mostly confined to the ingress and egress phases, as
might be expected given that it is defined as the change in specific intensity
due to limb darkening near the limb of the stars ($h_2 = I_X(\nicefrac{1}{2})
- I_X(0)$). It should be emphasized that changes in the transit depth and
shape due to the uncertainties in  $h_1$ and $h_2$ are very small ($\la 50$
ppm) and so will only be noticable in light curves of the very highest quality
for bright systems with deep eclipses. Indeed, it may be that some of the
offset between the observed and computed values of $h_1$ and $h_2$ is due to
small systematic errors in the photometry.  In general, the current level of
uncertainty in the computed values of  $h_1$ and $h_2$ will have a negligible
impact on the analysis of most light curves at optical wavelengths for many
transiting exoplanets. Fig.~\ref{h1h2lc} also demonstrates that using a sphere
to  optimize of the power-2 limb-darkening law coefficients instead of an
ellipsoid will have a negligible effect.

\subsection{Using the power-2 limb-darkening law}
 The analysis above supports the conclusion of \citet{2017AJ....154..111M}
that the power-2 limb-darkening law is to be recommended for the analysis of
transiting exoplanet light curves.  The atmospheric parameters (T$_{\rm
ref}$, $\log g$ and [Fe/H]) for many planet host stars are covered by the grid
of parameters for the  power-2 limb-darkening law provided in
Table~\ref{Power2Table}. The power-2 limb-darkening law has been implemented
in the light curve model {\sc ellc} and can also be used with other light
curve models, e.g., \texttt{batman} \citep{2015PASP..127.1161K}. It is quite
straightforward to implement this simple law in other light curve models.
There does not appear to be any reason why the  power-2 limb-darkening law
cannot also be used to model eclipsing binary stars, and the results in
Fig.~\ref{h1h2Teff} suggest that the convenient properties of the transformed
parameters $h_1$ and $h_2$ apply equally to these systems as for transiting
exoplanet systems.

The transformed parameters  $h_1$ and $h_2$ also make it straightforward to
calculate a Bayesian prior that accounts for the uncertainties in the model
coefficients when calculating the likelihood during  the analysis of light
curves using methods such as {\sc emcee} or other Markov chain methods. For
passbands similar to Kepler, e.g., CHEOPS, CoRoT and Gaia, the results from
the analysis above can be used directly. For stars that do not show strong
magnetic activity it can be assumed that $h_1$ is accurate to $\pm \sigma_1 =
\pm 0.011$ and $h_2$ is accurate to  $\pm \sigma_2 = \pm0.045$. The Bayesian
prior can then be calculated using a Gaussian centred on the values of $h_1$
and $h_2$ interpolated from Table~\ref{Power2Table} with standard deviations
$\sigma_{1, {\rm tot}} = \sqrt{\sigma_1 + \sigma_{1 ,{\rm obs}} }$, where
$\sigma_{1, {\rm obs}}$ is the standard error on $h_1$ due to the
uncertainties in the observed values of T$_{\rm eff}$, $\log g$ and [Fe/H],
and similarly for $\sigma_{2, {\rm obs}}$. The values of $\sigma_{1, {\rm
obs}}$ and $\sigma_{2, {\rm obs}}$ can be calculated using a Monte Carlo
method, as described in Section~\ref{p2sect}.

Limb darkening becomes less important at longer wavelengths so the same
assumptions can also be made for redder passbands such as TESS, Johnson I band
and Sloan $z^{\prime}$ band. This assumptions may be a little pessimistic, but
this is a sensible approach to take until the cause of the uncertainties in
$h_1$ and $h_2$ are better understood. For bandpasses such as Johnson B band
that cover shorter wavelengths where limb darkening is a stronger effect I
would recommend increasing the assumed values of $\sigma_1$ and $\sigma_2$ by
a factor of 2 or 3, perhaps even more for the Johnson U band and Sloan
$u^{\prime}$ band where the effect of uncertainties in the atomic data for the
large number of atomic lines in these strongly line-blanketed regions may be
much worse.

 The analysis above only quantifies the uncertainties in the model
coefficients for dwarf stars with  [Fe/H] $\ga 0$ showing weak magnetic
activity. It is likely that the model uncertainties are larger for stars that
are magnetically active and may well be biased in the sense that $h_1$ is too
low and $h_2$ is too high. However, this conclusion is only based on one
magnetically active star in the sample studied so far so it needs further
investigation. Similarly, there is currently very little information available
to quantify  the uncertainties in the  power-2 limb-darkening law coefficients
for metal poor stars or giant stars. 

\section{Conclusions}

 The power-2 limb-darkening law can be recommended for the analysis of light
curves for transiting exoplanet systems and binary stars for stars with
T$_{\rm eff}$, $\log g$ and [Fe/H] within the model grid range studied here.
Tabulations of the parameters of the power-2 limb-darkening law have been
provided and tested against very high quality observations of transiting
exoplanet systems obtained with Kepler. These observations have been used to
quantify the uncertainties in the parameters  $h_1$  and $h_2$ for dwarf stars
with [Fe/H] $\ga 0$ showing weak magnetic activity. There may be a small bias
in the computed values of $h_1$  and $h_2$ compared to the best-fit values for
magneticically active stars, but this needs further investigation. Further
work is also needed to quantify the uncertainties in $h_1$  and $h_2$ for
metal poor stars and red giants.

\begin{acknowledgements}
This paper includes data collected by the K2 mission. Funding for the K2
mission is provided by the NASA Science Mission directorate.

This research has made use of the SIMBAD database, operated at CDS,
Strasbourg, France. 

PM would like to thank Andrea Chiavassa and Martin Asplund for providing
additional limb-darkening data calculated from the {\sc Stagger}-grid.

\end{acknowledgements}

\bibliographystyle{aa} 
\bibliography{mybib}

\begin{thebibliography}{40}
\expandafter\ifx\csname natexlab\endcsname\relax\def\natexlab#1{#1}\fi

\bibitem[{{Auvergne} {et~al.}(2009)}]{2009A&A...506..411A}
{Auvergne}, M. {et~al.} 2009, \aap, 506, 411

\bibitem[{{Bessell}(1990)}]{1990PASP..102.1181B}
{Bessell}, M.~S. 1990, \pasp, 102, 1181

\bibitem[{{Bessell} {et~al.}(1998){Bessell}, {Castelli}, \&
  {Plez}}]{1998A+A...333..231B}
{Bessell}, M.~S., {Castelli}, F., \& {Plez}, B. 1998, \aap, 333, 231

\bibitem[{{Bigot} {et~al.}(2006){Bigot}, {Kervella}, {Th{\'e}venin}, \&
  {S{\'e}gransan}}]{2006A&A...446..635B}
{Bigot}, L., {Kervella}, P., {Th{\'e}venin}, F., \& {S{\'e}gransan}, D. 2006,
  \aap, 446, 635

\bibitem[{{Borucki} {et~al.}(2010){Borucki}, {Koch}, {Basri}, {Batalha},
  {Brown}, {Caldwell}, {Caldwell}, {Christensen-Dalsgaard}, {Cochran},
  {DeVore}, {Dunham}, {Dupree}, {Gautier}, {Geary}, {Gilliland}, {Gould},
  {Howell}, {Jenkins}, {Kondo}, {Latham}, {Marcy}, {Meibom}, {Kjeldsen},
  {Lissauer}, {Monet}, {Morrison}, {Sasselov}, {Tarter}, {Boss}, {Brownlee},
  {Owen}, {Buzasi}, {Charbonneau}, {Doyle}, {Fortney}, {Ford}, {Holman},
  {Seager}, {Steffen}, {Welsh}, {Rowe}, {Anderson}, {Buchhave}, {Ciardi},
  {Walkowicz}, {Sherry}, {Horch}, {Isaacson}, {Everett}, {Fischer}, {Torres},
  {Johnson}, {Endl}, {MacQueen}, {Bryson}, {Dotson}, {Haas}, {Kolodziejczak},
  {Van Cleve}, {Chandrasekaran}, {Twicken}, {Quintana}, {Clarke}, {Allen},
  {Li}, {Wu}, {Tenenbaum}, {Verner}, {Bruhweiler}, {Barnes}, \&
  {Prsa}}]{2010Sci...327..977B}
{Borucki}, W.~J., {Koch}, D., {Basri}, G., {et~al.} 2010, Science, 327, 977

\bibitem[{{Cessa} {et~al.}(2017){Cessa}, {Beck}, {Benz}, {Broeg}, {Ehrenreich},
  {Fortier}, {Peter}, {Magrin}, {Pagano}, {Plesseria}, {Steller}, {Szoke},
  {Thomas}, {Ragazzoni}, \& {Wildi}}]{2017SPIE10563E..1LC}
{Cessa}, V., {Beck}, T., {Benz}, W., {et~al.} 2017, in Society of Photo-Optical
  Instrumentation Engineers (SPIE) Conference Series, Vol. 10563, Society of
  Photo-Optical Instrumentation Engineers (SPIE) Conference Series, 105631L

\bibitem[{{Chiavassa} {et~al.}(2018{\natexlab{a}}){Chiavassa}, {Casagrande},
  {Collet}, {Magic}, {Bigot}, {Thevenin}, \& {Asplund}}]{2018arXiv180101895C}
{Chiavassa}, A., {Casagrande}, L., {Collet}, R., {et~al.} 2018{\natexlab{a}},
  ArXiv e-prints

\bibitem[{{Chiavassa} {et~al.}(2018{\natexlab{b}}){Chiavassa}, {Casagrande},
  {Collet}, {Magic}, {Bigot}, {Th{\'e}venin}, \&
  {Asplund}}]{2018A&A...611A..11C}
{Chiavassa}, A., {Casagrande}, L., {Collet}, R., {et~al.} 2018{\natexlab{b}},
  \aap, 611, A11

\bibitem[{{Claret}(2000)}]{2000A&A...363.1081C}
{Claret}, A. 2000, \aap, 363, 1081

\bibitem[{{Claret}(2004)}]{2004A&A...428.1001C}
{Claret}, A. 2004, \aap, 428, 1001

\bibitem[{{Claret} {et~al.}(2013){Claret}, {Hauschildt}, \&
  {Witte}}]{2013A&A...552A..16C}
{Claret}, A., {Hauschildt}, P.~H., \& {Witte}, S. 2013, \aap, 552, A16

\bibitem[{{Doi} {et~al.}(2010){Doi}, {Tanaka}, {Fukugita}, {Gunn}, {Yasuda},
  {Ivezić}, {Brinkmann}, {de Haars}, {Kleinman}, {Krzesinski}, \& {French
  Leger}}]{2010AJ....139.1628D}
{Doi}, M., {Tanaka}, M., {Fukugita}, M., {et~al.} 2010, \aj, 139, 1628

\bibitem[{{Espinoza} \& {Jord{\'a}n}(2016)}]{2016MNRAS.457.3573E}
{Espinoza}, N. \& {Jord{\'a}n}, A. 2016, \mnras, 457, 3573

\bibitem[{{Foreman-Mackey} {et~al.}(2017){Foreman-Mackey}, {Agol},
  {Ambikasaran}, \& {Angus}}]{2017AJ....154..220F}
{Foreman-Mackey}, D., {Agol}, E., {Ambikasaran}, S., \& {Angus}, R. 2017, \aj,
  154, 220

\bibitem[{{Foreman-Mackey} {et~al.}(2013){Foreman-Mackey}, {Hogg}, {Lang}, \&
  {Goodman}}]{2013PASP..125..306F}
{Foreman-Mackey}, D., {Hogg}, D.~W., {Lang}, D., \& {Goodman}, J. 2013, \pasp,
  125, 306

\bibitem[{{Gaia Collaboration} {et~al.}(2016){Gaia Collaboration}, {Prusti},
  {de Bruijne}, {Brown}, {Vallenari}, {Babusiaux}, {Bailer-Jones}, {Bastian},
  {Biermann}, {Evans}, \& et~al.}]{2016A&A...595A...1G}
{Gaia Collaboration}, {Prusti}, T., {de Bruijne}, J.~H.~J., {et~al.} 2016,
  \aap, 595, A1

\bibitem[{{Hayek} {et~al.}(2012){Hayek}, {Sing}, {Pont}, \&
  {Asplund}}]{2012A&A...539A.102H}
{Hayek}, W., {Sing}, D., {Pont}, F., \& {Asplund}, M. 2012, \aap, 539, A102

\bibitem[{{H{\'e}brard} {et~al.}(2014){H{\'e}brard}, {Santerne}, {Montagnier},
  {Bruno}, {Deleuil}, {Havel}, {Almenara}, {Damiani}, {Barros}, {Bonomo},
  {Bouchy}, {D{\'{\i}}az}, \& {Moutou}}]{2014A&A...572A..93H}
{H{\'e}brard}, G., {Santerne}, A., {Montagnier}, G., {et~al.} 2014, \aap, 572,
  A93

\bibitem[{{Hestroffer}(1997)}]{1997A&A...327..199H}
{Hestroffer}, D. 1997, \aap, 327, 199

\bibitem[{{Howarth}(2011)}]{2011MNRAS.418.1165H}
{Howarth}, I.~D. 2011, \mnras, 418, 1165

\bibitem[{{Kipping} \& {Bakos}(2011)}]{2011ApJ...730...50K}
{Kipping}, D. \& {Bakos}, G. 2011, \apj, 730, 50

\bibitem[{{Kipping}(2013)}]{2013MNRAS.435.2152K}
{Kipping}, D.~M. 2013, \mnras, 435, 2152

\bibitem[{{Knutson} {et~al.}(2007){Knutson}, {Charbonneau}, {Noyes}, {Brown},
  \& {Gilliland}}]{2007ApJ...655..564K}
{Knutson}, H.~A., {Charbonneau}, D., {Noyes}, R.~W., {Brown}, T.~M., \&
  {Gilliland}, R.~L. 2007, \apj, 655, 564

\bibitem[{{Kopal}(1950)}]{1950HarCi.454....1K}
{Kopal}, Z. 1950, Harvard College Observatory Circular, 454, 1

\bibitem[{{Kreidberg}(2015)}]{2015PASP..127.1161K}
{Kreidberg}, L. 2015, \pasp, 127, 1161

\bibitem[{{Magic} {et~al.}(2015){Magic}, {Chiavassa}, {Collet}, \&
  {Asplund}}]{2015A&A...573A..90M}
{Magic}, Z., {Chiavassa}, A., {Collet}, R., \& {Asplund}, M. 2015, \aap, 573,
  A90

\bibitem[{{Maxted}(2016)}]{2016A&A...591A.111M}
{Maxted}, P.~F.~L. 2016, \aap, 591, A111

\bibitem[{{Maxted} {et~al.}(2016){Maxted}, {Anderson}, {Collier Cameron},
  {Delrez}, {Gillon}, {Hellier}, {Jehin}, {Lendl}, {Neveu-VanMalle}, {Pepe},
  {Pollacco}, {Queloz}, {S{\'e}gransan}, {Smalley}, {Smith}, {Southworth},
  {Triaud}, {Udry}, {Wagg}, \& {West}}]{2016A&A...591A..55M}
{Maxted}, P.~F.~L., {Anderson}, D.~R., {Collier Cameron}, A., {et~al.} 2016,
  \aap, 591, A55

\bibitem[{{Morello} {et~al.}(2017){Morello}, {Tsiaras}, {Howarth}, \&
  {Homeier}}]{2017AJ....154..111M}
{Morello}, G., {Tsiaras}, A., {Howarth}, I.~D., \& {Homeier}, D. 2017, \aj,
  154, 111

\bibitem[{{Müller} {et~al.}(2013){Müller}, {Huber}, {Czesla}, {Wolter}, \&
  {Schmitt}}]{2013A&A...560A.112M}
{Müller}, H.~M., {Huber}, K.~F., {Czesla}, S., {Wolter}, U., \& {Schmitt},
  J.~H.~M.~M. 2013, \aap, 560, A112

\bibitem[{{Neilson} {et~al.}(2017){Neilson}, {McNeil}, {Ignace}, \&
  {Lester}}]{2017ApJ...845...65N}
{Neilson}, H.~R., {McNeil}, J.~T., {Ignace}, R., \& {Lester}, J.~B. 2017, \apj,
  845, 65

\bibitem[{{P{\'a}l}(2008)}]{2008MNRAS.390..281P}
{P{\'a}l}, A. 2008, \mnras, 390, 281

\bibitem[{{Pereira} {et~al.}(2013){Pereira}, {Asplund}, {Collet}, {Thaler},
  {Trampedach}, \& {Leenaarts}}]{2013A&A...554A.118P}
{Pereira}, T.~M.~D., {Asplund}, M., {Collet}, R., {et~al.} 2013, \aap, 554,
  A118

\bibitem[{{Ricker} {et~al.}(2015){Ricker}, {Winn}, {Vanderspek}, {Latham},
  {Bakos}, {Bean}, {Berta-Thompson}, {Brown}, {Buchhave}, {Butler}, {Butler},
  {Chaplin}, {Charbonneau}, {Christensen-Dalsgaard}, {Clampin}, {Deming},
  {Doty}, {De Lee}, {Dressing}, {Dunham}, {Endl}, {Fressin}, {Ge}, {Henning},
  {Holman}, {Howard}, {Ida}, {Jenkins}, {Jernigan}, {Johnson}, {Kaltenegger},
  {Kawai}, {Kjeldsen}, {Laughlin}, {Levine}, {Lin}, {Lissauer}, {MacQueen},
  {Marcy}, {McCullough}, {Morton}, {Narita}, {Paegert}, {Palle}, {Pepe},
  {Pepper}, {Quirrenbach}, {Rinehart}, {Sasselov}, {Sato}, {Seager},
  {Sozzetti}, {Stassun}, {Sullivan}, {Szentgyorgyi}, {Torres}, {Udry}, \&
  {Villasenor}}]{2015JATIS...1a4003R}
{Ricker}, G.~R., {Winn}, J.~N., {Vanderspek}, R., {et~al.} 2015, Journal of
  Astronomical Telescopes, Instruments, and Systems, 1, 014003

\bibitem[{{Santerne} {et~al.}(2012){Santerne}, {D{\'{\i}}az}, {Moutou},
  {Bouchy}, {H{\'e}brard}, {Almenara}, {Bonomo}, {Deleuil}, \&
  {Santos}}]{2012A&A...545A..76S}
{Santerne}, A., {D{\'{\i}}az}, R.~F., {Moutou}, C., {et~al.} 2012, \aap, 545,
  A76

\bibitem[{{Schwarzschild}(1906)}]{Schwarzschild1906}
{Schwarzschild}, K. 1906, Nachrichten von der Gesellschaft der Wissenschaften
  zu Göttingen, 43

\bibitem[{{Sing} {et~al.}(2008){Sing}, {Vidal-Madjar}, {D{\'e}sert},
  {Lecavelier des Etangs}, \& {Ballester}}]{2008ApJ...686..658S}
{Sing}, D.~K., {Vidal-Madjar}, A., {D{\'e}sert}, J.-M., {Lecavelier des
  Etangs}, A., \& {Ballester}, G. 2008, \apj, 686, 658

\bibitem[{{Southworth}(2011)}]{2011MNRAS.417.2166S}
{Southworth}, J. 2011, \mnras, 417, 2166

\bibitem[{{Thompson} {et~al.}(2016){Thompson}, {Fraquelli}, {Van Cleve}, \&
  {Caldwell}}]{2016ksci.rept....9T}
{Thompson}, S.~E., {Fraquelli}, D., {Van Cleve}, J.~E., \& {Caldwell}, D.~A.
  2016, {Kepler Archive Manual}, Tech. rep., Space Telescope Science Institute

\bibitem[{{Walker} {et~al.}(2003){Walker}, {Matthews}, {Kuschnig}, {Johnson},
  {Rucinski}, {Pazder}, {Burley}, {Walker}, {Skaret}, {Zee}, {Grocott},
  {Carroll}, {Sinclair}, {Sturgeon}, \& {Harron}}]{2003PASP..115.1023W}
{Walker}, G., {Matthews}, J., {Kuschnig}, R., {et~al.} 2003, \pasp, 115, 1023

\end{thebibliography}

\appendix

\section{Light curve analysis -- results}
  The results for the fits to the light curves of 16 transiting
exoplanet systems are shown in this appendix.

\begin{table*}
\caption{Results for \texttt{ellc} light curve model fits to Kepler light
curves of transiting exoplanet systems. The values of T$_{\rm
eff}$, $\log g$ and [Fe/H] are taken from TEPCat unless otherwise noted in the
footnotes to this table. The values and error quoted for $h_{1,{\rm obs}}$,
  $h_{2,{\rm obs}}$, $R_{\star}/a$, $k$ and $b$ are the median and standard
deviation of the posterior probability distributions calculated using {\sc
emcee}. The number of data points used in the analysis, N$_{\rm fit}$,  and the
standard deviation of the residuals from the best fit, $\sigma_{\rm res}$, are
noted in the final column.}
\label{ResultsTable}
\begin{center}
\begin{tabular}{@{}lrrrrrrrrrrrr}
\hline
  \multicolumn{1}{@{}l}{Name} &
  \multicolumn{1}{c}{T$_{\rm eff}$ [K]} &
  \multicolumn{1}{c}{$\log g$ [cgs]} &
  \multicolumn{1}{c}{[Fe/H]} &
  \multicolumn{1}{c}{$h_{\rm 1,comp}$} &
  \multicolumn{1}{c}{$h_{\rm 1,obs}$} &
  \multicolumn{1}{c}{$h_{\rm 2,comp}$} &
  \multicolumn{1}{c}{$h_{\rm 2,obs}$} &
  \multicolumn{1}{c}{$R_{\star}/a$} &
  \multicolumn{1}{c}{$k$} &
  \multicolumn{1}{c}{$b$} &
  \multicolumn{1}{c}{N$_{\rm fit}$}   \\
  \multicolumn{1}{@{}l}{KIC} &
  \multicolumn{1}{c}{$\pm$}&
  \multicolumn{1}{c}{$\pm$}&
  \multicolumn{1}{c}{$\pm$}&
  \multicolumn{1}{c}{$\pm$}&
  \multicolumn{1}{c}{$\pm$}&
  \multicolumn{1}{c}{$\pm$}&
  \multicolumn{1}{c}{$\pm$}&
  \multicolumn{1}{c}{$\pm$}&
  \multicolumn{1}{c}{$\pm$}&
  \multicolumn{1}{c}{$\pm$}&
  \multicolumn{1}{l}{$\sigma_{\rm res}$}   \\
\hline
\noalign{\smallskip}
Kepler-5  & 6297& 4.17 &$ +0.00$& $0.763$&$ 0.788$&$ 0.456$&$ 0.37$ & 0.15439& 0.07915& 0.034 &   38852\\
  8191672 &   60& 0.02 &   0.00 & $0.002$&$ 0.002$&$ 0.001$&$ 0.04$ & 0.00024& 0.00016& 0.028 & 0.00067\\
\noalign{\smallskip}
Kepler-6  & 5647& 4.28 &$ +0.34$& $0.718$&$ 0.731$&$ 0.444$&$ 0.34$ & 0.13161& 0.09349& 0.021 &   59279\\
 10874614 &   50& 0.02 &   0.05 & $0.004$&$ 0.001$&$ 0.002$&$ 0.02$ & 0.00009& 0.00005& 0.016 & 0.00062\\
\noalign{\smallskip}
Kepler-7  & 5933& 3.97 &$ +0.11$& $0.747$&$ 0.762$&$ 0.444$&$ 0.45$ & 0.14931& 0.08248& 0.548 &   41449\\
  5780885 &   50& 0.02 &   0.05 & $0.002$&$ 0.002$&$ 0.003$&$ 0.04$ & 0.00049& 0.00012& 0.005 & 0.00049\\
\noalign{\smallskip}
Kepler-8  & 6213& 4.18 &$ -0.06$& $0.759$&$ 0.772$&$ 0.450$&$ 0.40$ & 0.14527& 0.09490& 0.715 &   68349\\
  6922244 &  150& 0.02 &   0.05 & $0.008$&$ 0.004$&$ 0.001$&$ 0.05$ & 0.00044& 0.00015& 0.003 & 0.00092\\
\noalign{\smallskip}
Kepler-12 & 5947& 4.16 &$ +0.07$& $0.743$&$ 0.757$&$ 0.457$&$ 0.42$ & 0.12365& 0.11800& 0.102 &   59828\\
 11804465 &  100& 0.02 &   0.04 & $0.005$&$ 0.001$&$ 0.001$&$ 0.03$ & 0.00018& 0.00010& 0.019 & 0.00076\\
\noalign{\smallskip}
Kepler-15 & 5595& 4.28 &$ +0.36$& $0.714$&$ 0.721$&$ 0.443$&$ 0.39$ & 0.10052& 0.10298& 0.676 &   19982\\
 11359879 &  120& 0.02 &   0.07 & $0.009$&$ 0.004$&$ 0.005$&$ 0.05$ & 0.00040& 0.00025& 0.004 & 0.00081\\
\noalign{\smallskip}
Kepler-17 & 5595& 4.28 &$ +0.36$& $0.714$&$ 0.770$&$ 0.445$&$ 0.21$ & 0.17316& 0.13350& 0.012 &  104916\\
 10619192 &  120& 0.02 &   0.10 & $0.007$&$ 0.001$&$ 0.005$&$ 0.01$ & 0.00010& 0.00010& 0.010 & 0.00138\\
\noalign{\smallskip}
Kepler-41 & 5750& 4.28 &$ +0.38$& $0.726$&$ 0.726$&$ 0.443$&$ 0.41$ & 0.19298& 0.10109& 0.669 &   29921\\
  9410930 &  100& 0.01 &   0.11 & $0.006$&$ 0.006$&$ 0.006$&$ 0.06$ & 0.00102& 0.00030& 0.006 & 0.00121\\
\noalign{\smallskip}
Kepler-43 & 6041& 4.28 &$ +0.33$& $0.741$&$ 0.750$&$ 0.453$&$ 0.44$ & 0.14157& 0.08541& 0.641 &   65553\\
  9818381 &  143& 0.02 &   0.11 & $0.009$&$ 0.003$&$ 0.005$&$ 0.05$ & 0.00060& 0.00017& 0.005 & 0.00086\\
\noalign{\smallskip}
Kepler-77 & 5520& 4.42 &$ +0.20$& $0.712$&$ 0.718$&$ 0.442$&$ 0.43$ & 0.10056& 0.09743& 0.282 &   24419\\
  8359498 &   60& 0.01 &   0.05 & $0.004$&$ 0.003$&$ 0.003$&$ 0.06$ & 0.00064& 0.00031& 0.024 & 0.00094\\
\noalign{\smallskip}
Kepler-412 & 5750& 4.30 &$ +0.27$& $0.726$&$ 0.738$&$ 0.442$&$ 0.44$ & 0.20219& 0.10519& 0.786 &   28134\\
  7877496 &   90& 0.07 &   0.12 & $0.006$&$ 0.010$&$ 0.005$&$ 0.08$ & 0.00110& 0.00061& 0.004 & 0.00110\\
\noalign{\smallskip}
Kepler-422 & 5972& 4.31 &$ +0.23$& $0.737$&$ 0.741$&$ 0.450$&$ 0.41$ & 0.07280& 0.09554& 0.475 &   43453\\
  9631995 &   84& 0.07 &   0.09 & $0.006$&$ 0.002$&$ 0.004$&$ 0.04$ & 0.00025& 0.00016& 0.007 & 0.00074\\
\noalign{\smallskip}
Kepler-423 & 5560& 4.41 &$ -0.10$& $0.722$&$ 0.735$&$ 0.451$&$ 0.40$ & 0.12040& 0.12417& 0.268 &   41034\\
  9651668 &   80& 0.04 &   0.05 & $0.005$&$ 0.002$&$ 0.001$&$ 0.05$ & 0.00043& 0.00032& 0.016 & 0.00113\\
\noalign{\smallskip}
Kepler-425 & 5170& 4.54 &$ +0.24$& $0.689$&$ 0.697$&$ 0.435$&$ 0.40$ & 0.08447& 0.11411& 0.581 &   14339\\
  5357901 &   70& 0.01 &   0.11 & $0.004$&$ 0.005$&$ 0.005$&$ 0.07$ & 0.00061& 0.00051& 0.011 & 0.00132\\
\noalign{\smallskip}
Kepler-491 & 5634& 4.37 &$ +0.42$& $0.713$&$ 0.710$&$ 0.435$&$ 0.43$ & 0.08717& 0.08003& 0.384 &   15078\\
  6849046 &  114& 0.11 &   0.14 & $0.007$&$ 0.004$&$ 0.004$&$ 0.07$ & 0.00118& 0.00044& 0.036 & 0.00087\\
\noalign{\smallskip}
TrES-2    & 5850& 4.47 &$ -0.15$& $0.738$&$ 0.756$&$ 0.451$&$ 0.33$ & 0.12555& 0.12612& 0.845 &   13726\\
 11446443 &   50& 0.01 &   0.10 & $0.003$&$ 0.011$&$ 0.001$&$ 0.03$ & 0.00027& 0.00034& 0.001 & 0.00024\\
\noalign{\smallskip}
\hline
\end{tabular}
\tablefoot{
  \tablefoottext{a}{\citet{2014A&A...572A..93H}}
  \tablefoottext{b}{\citet{2012A&A...545A..76S}}
}
\end{center}
\end{table*}

\begin{figure*}
  \includegraphics[width=0.49\textwidth]{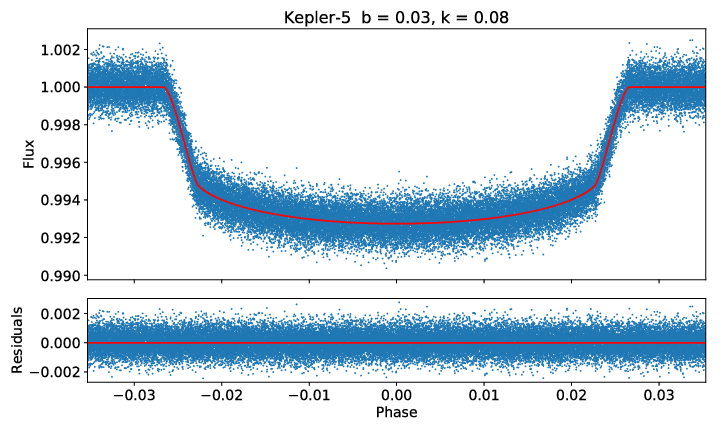}
  \includegraphics[width=0.49\textwidth]{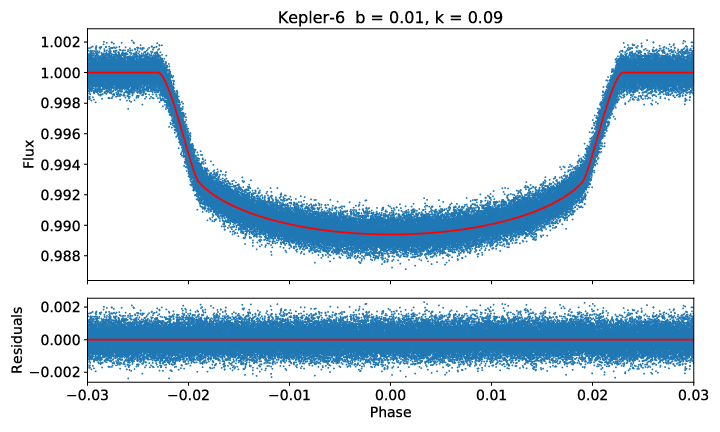}
  \includegraphics[width=0.49\textwidth]{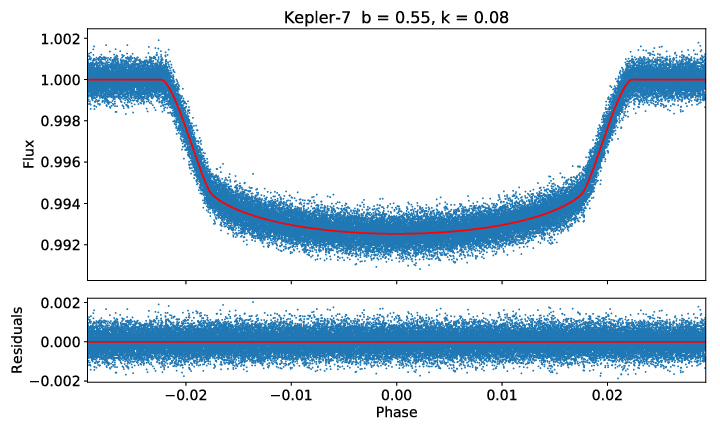}
  \includegraphics[width=0.49\textwidth]{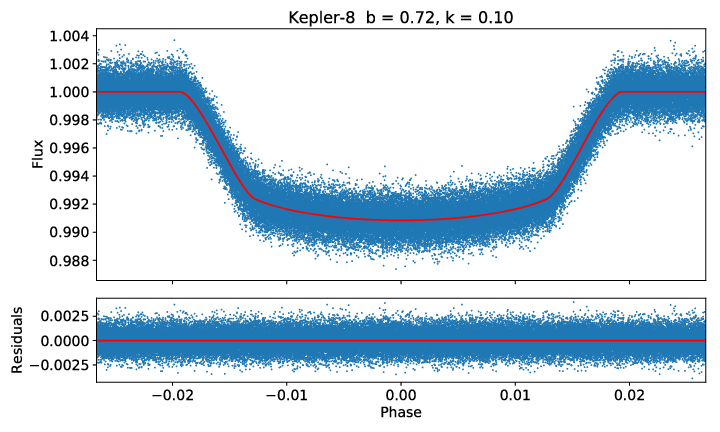}
  \includegraphics[width=0.49\textwidth]{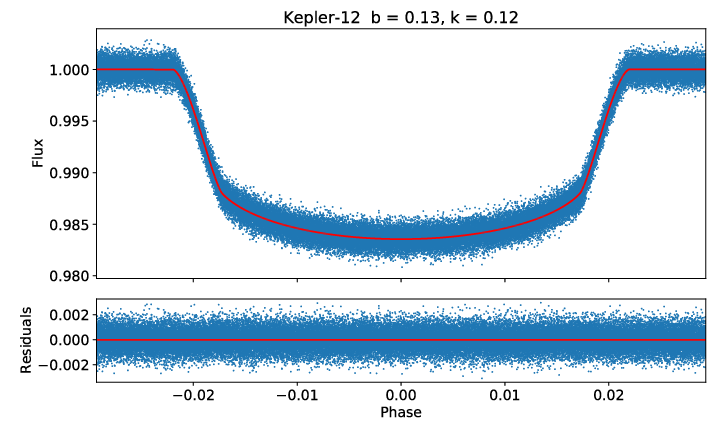}
  \includegraphics[width=0.49\textwidth]{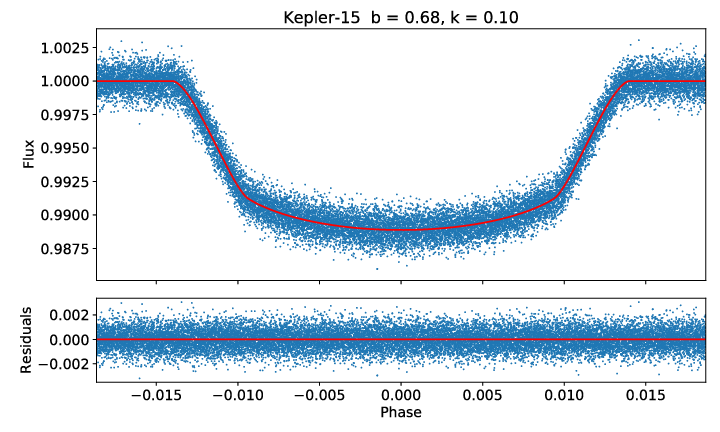}
  \includegraphics[width=0.49\textwidth]{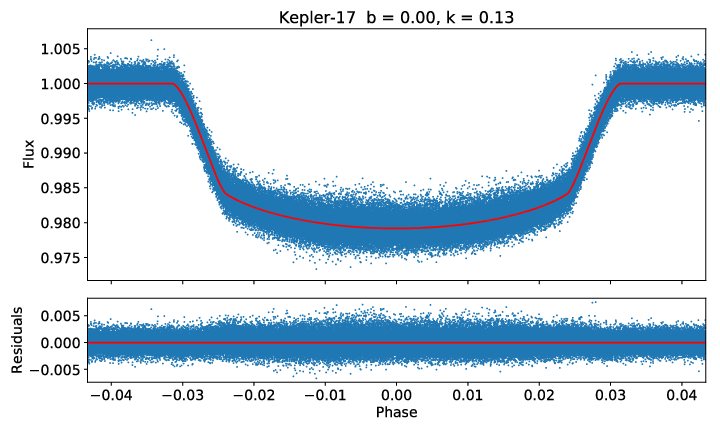}
  \includegraphics[width=0.49\textwidth]{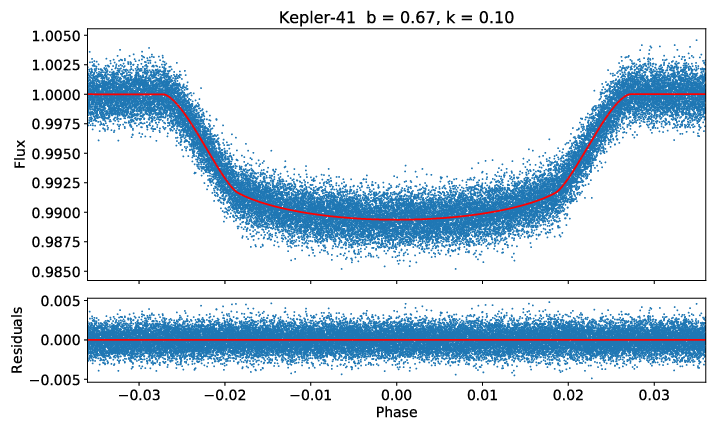}
\caption{Kepler light curves of transiting exoplanet and binary star systems.
  Observations are plotted using small points and the best-fit light curve
  model is shown as a line. The name of each star is noted in the title to
  each panel together with the impact parameter $b=a\cos{i}/R_{\star}$ and the
  ratio of the radii $k=R_{pl}/R_{\star}$.}
  \label{lcfitFig}
\end{figure*}

\begin{figure*}
\addtocounter{figure}{-1}
  \includegraphics[width=0.49\textwidth]{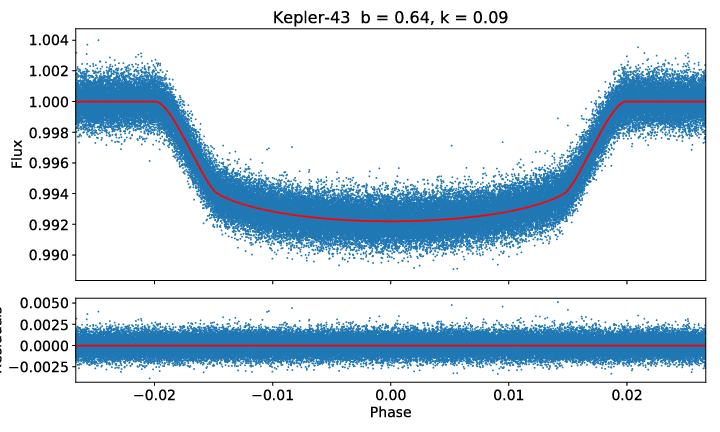}
  \includegraphics[width=0.49\textwidth]{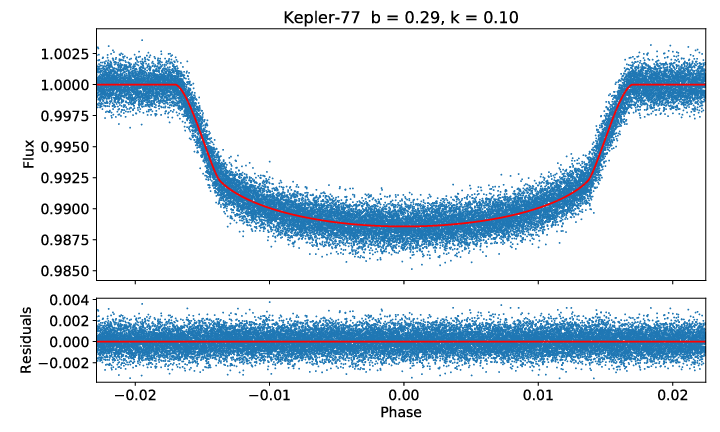}
  \includegraphics[width=0.49\textwidth]{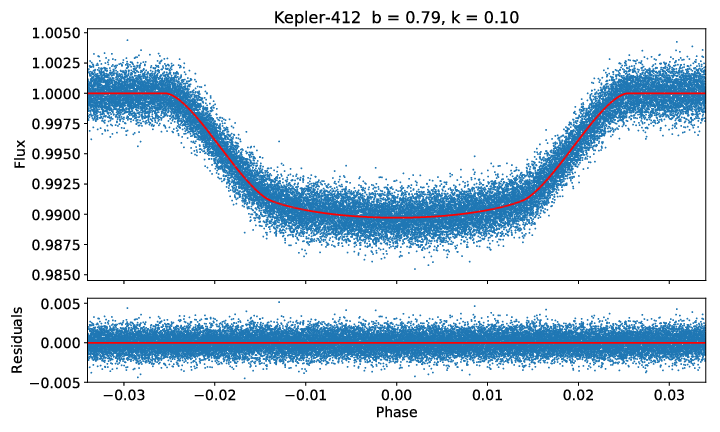}
  \includegraphics[width=0.49\textwidth]{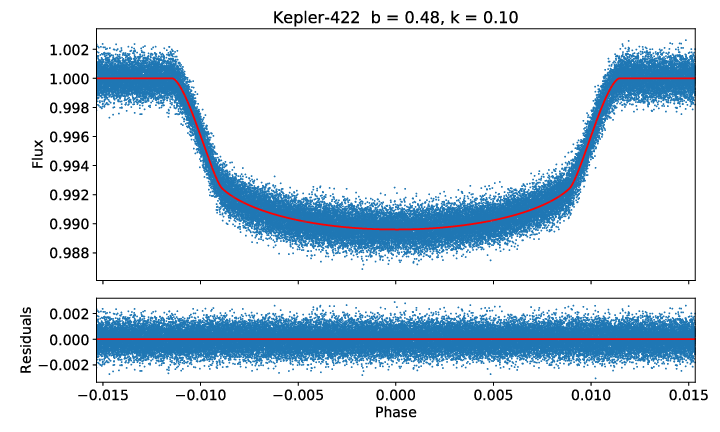}
  \includegraphics[width=0.49\textwidth]{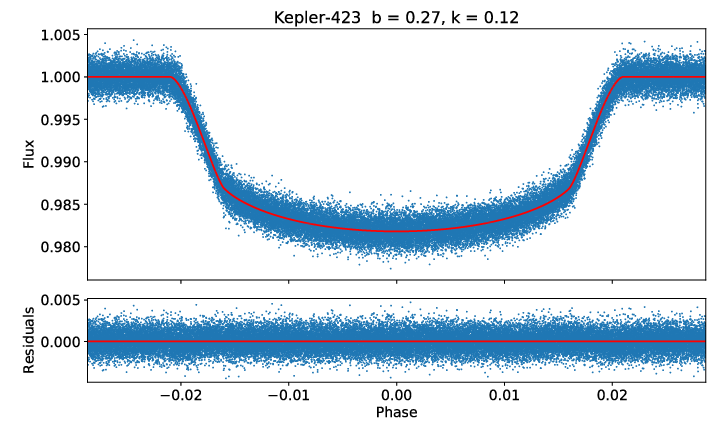}
  \includegraphics[width=0.49\textwidth]{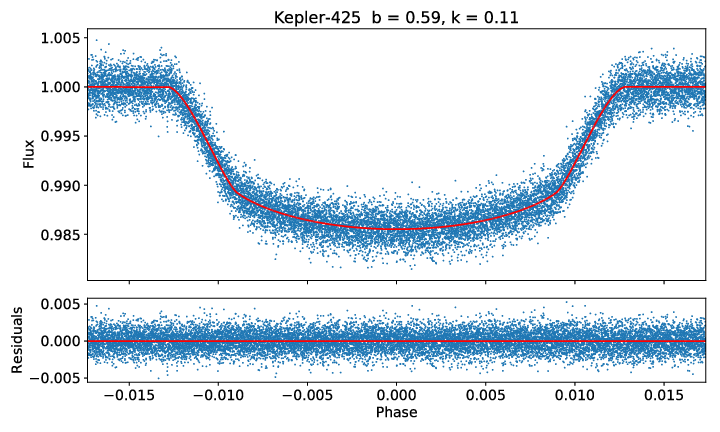}
  \includegraphics[width=0.49\textwidth]{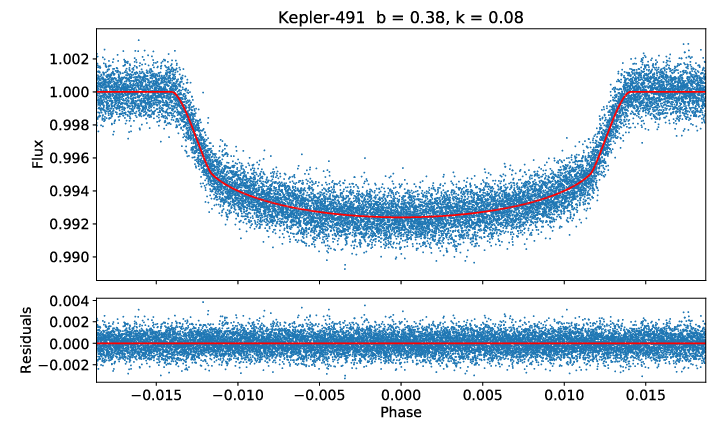}
  \includegraphics[width=0.49\textwidth]{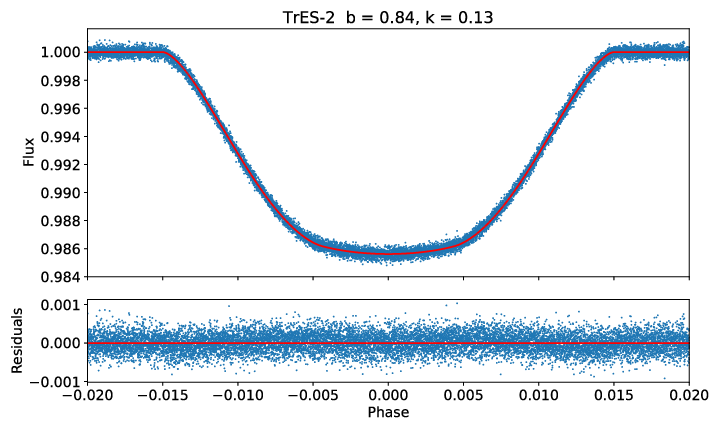}
  \caption{ continued.}
\end{figure*}

\begin{figure*}
  \includegraphics[width=0.49\textwidth]{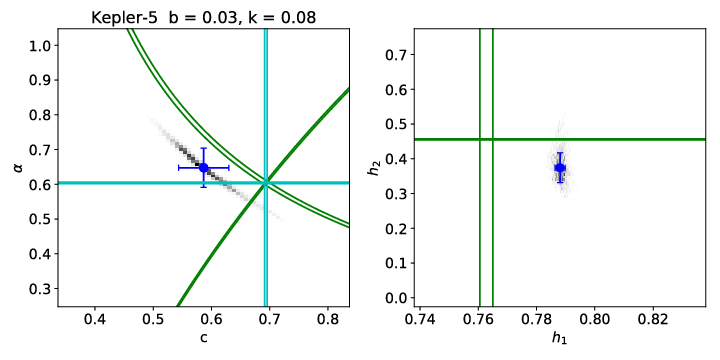}
  \includegraphics[width=0.49\textwidth]{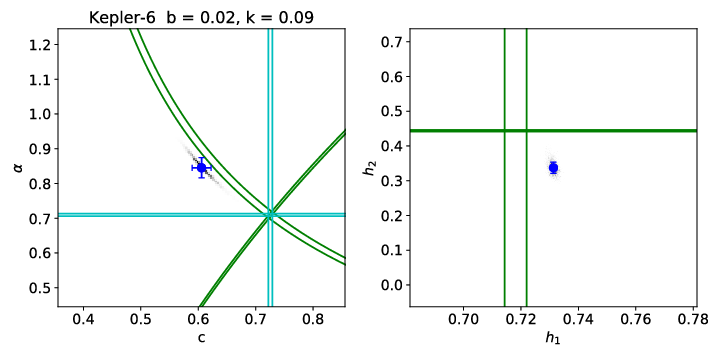}
  \includegraphics[width=0.49\textwidth]{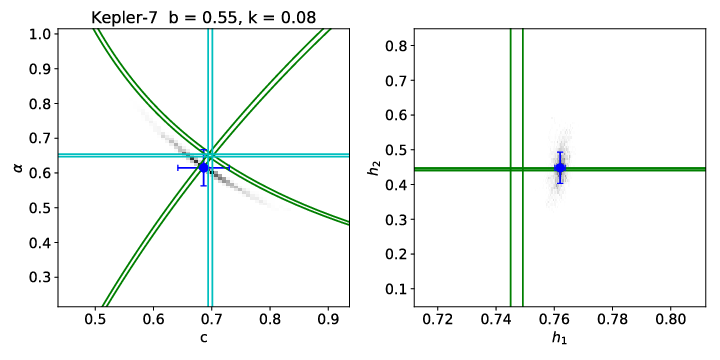}
  \includegraphics[width=0.49\textwidth]{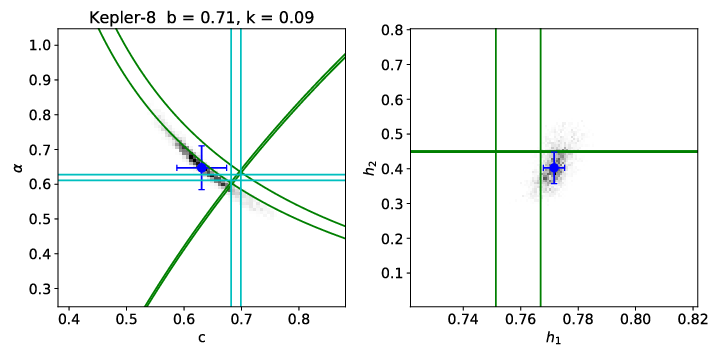}
  \includegraphics[width=0.49\textwidth]{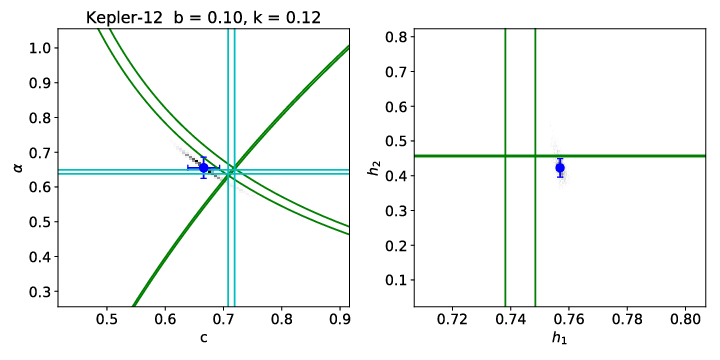}
  \includegraphics[width=0.49\textwidth]{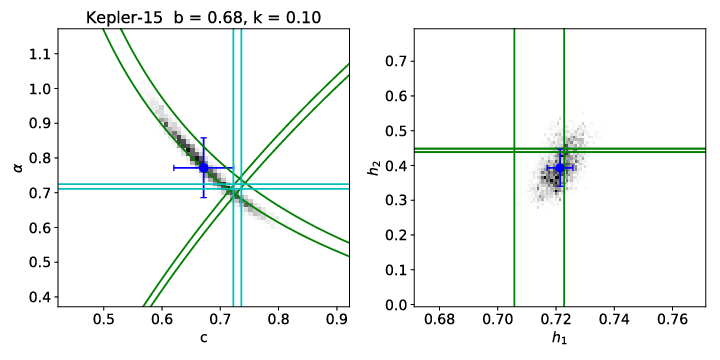}
  \includegraphics[width=0.49\textwidth]{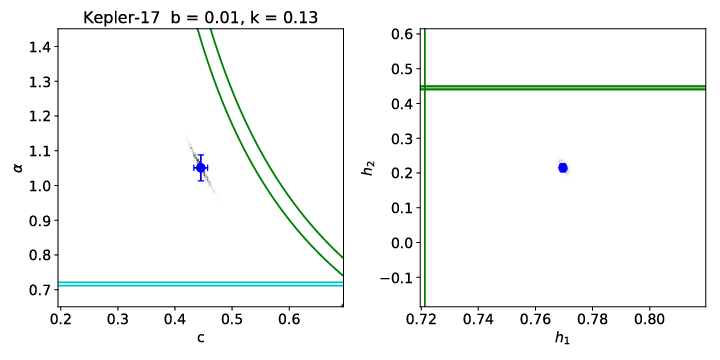}
  \includegraphics[width=0.49\textwidth]{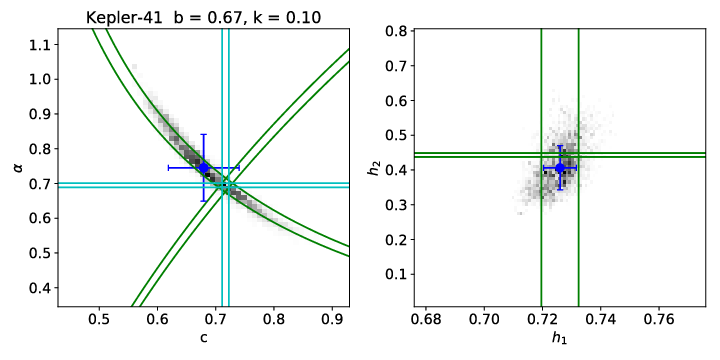}
  \caption{Posterior probability distributions for the limb-darkening
  parameters from the analysis of the Kepler light curves. The PPDs are
  shown as gray-scale density plots. The median values and standard deviations
  for each parameter are shown as an error bar.  The vertical and horizontal
  lines in the left-hand panel for each star show the $\pm 1$-$\sigma$ limits
  on the computed values of $c$ and $\alpha$. The corresponding  $\pm
  1$-$\sigma$ limits on $h_1$ and $h_2$ are shown as curved lines in the same
  panel and as vertical and horizontal lines in the right-hand panel. }
  \label{h1h2Fig}
\end{figure*}

\begin{figure*}
\addtocounter{figure}{-1}
  \includegraphics[width=0.49\textwidth]{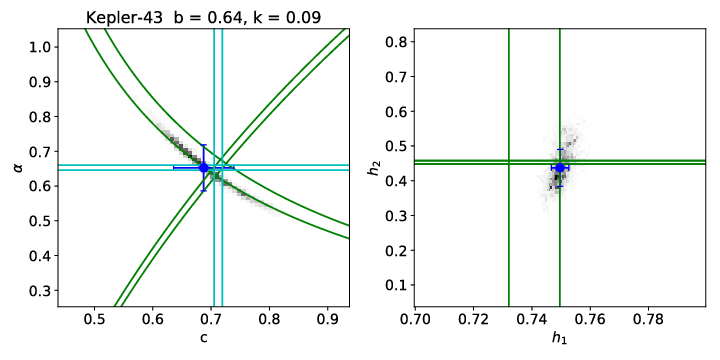}
  \includegraphics[width=0.49\textwidth]{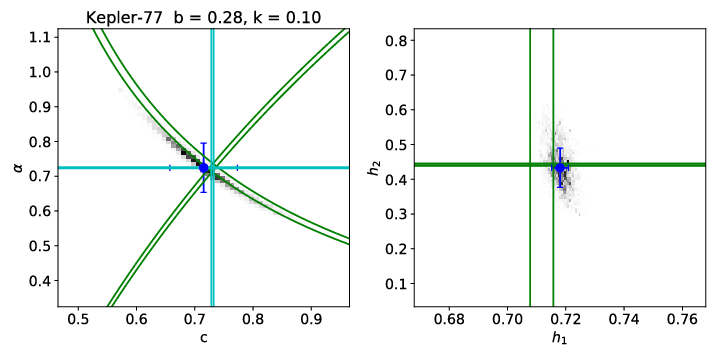}
  \includegraphics[width=0.49\textwidth]{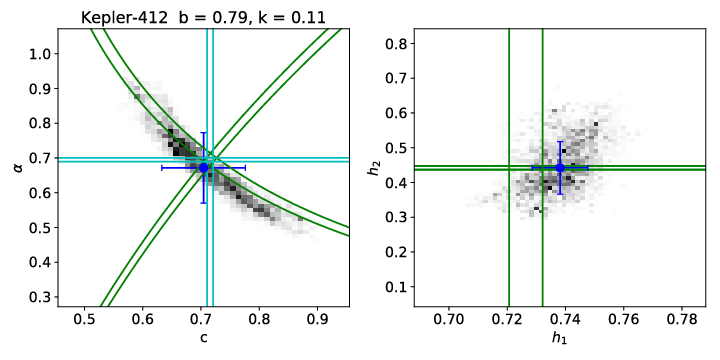}
  \includegraphics[width=0.49\textwidth]{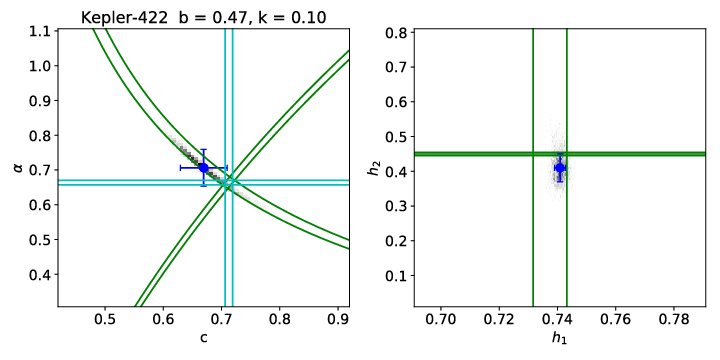}
  \includegraphics[width=0.49\textwidth]{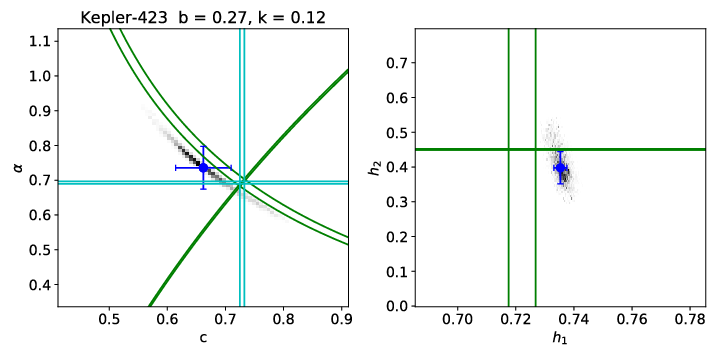}
  \includegraphics[width=0.49\textwidth]{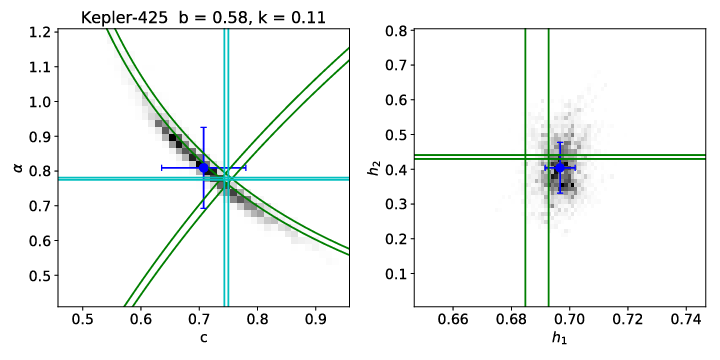}
  \includegraphics[width=0.49\textwidth]{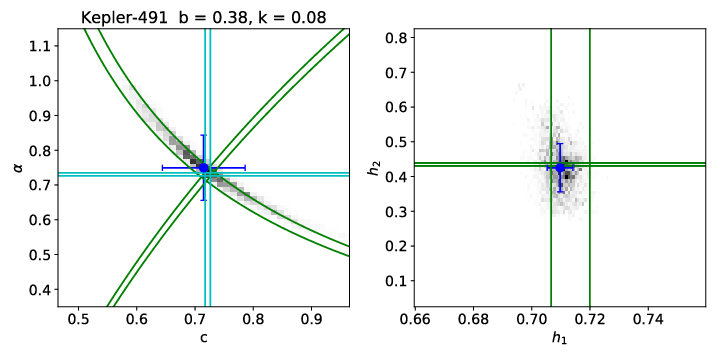}
  \includegraphics[width=0.49\textwidth]{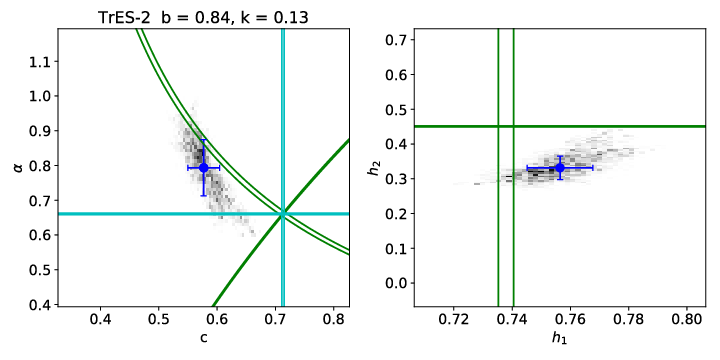}
  \caption{ continued.}
\end{figure*}

\end{document}